\documentclass[preprint,12pt]{elsarticle}
\usepackage{amssymb}
\usepackage[hmargin=3cm,top=3cm,bottom=3cm]{geometry}
\usepackage[nottoc,notlot,notlof]{tocbibind}
\usepackage[graphicx]{realboxes}
\graphicspath{{images/}}
\usepackage{hyperref}
\hypersetup{
    citecolor=blue,
    linkcolor=red,
    colorlinks=TRUE,
    pdftitle={How Does China's Household Portfolio Selection Vary with Financial Inclusion?},
    pdfpagemode=FullScreen,
}
\usepackage[capposition=top]{floatrow}
\usepackage{float}
\usepackage{setspace}
\usepackage{pdflscape}
\usepackage{xurl}
\usepackage[mathlines,displaymath]{lineno}
\usepackage{amsmath,nccmath}
\usepackage{enumitem}
\usepackage{makecell}
\usepackage{subcaption}
\usepackage[skip=3pt]{caption}
\usepackage{appendix}
\usepackage{array}
\usepackage{threeparttablex}
\usepackage{rotating}
\usepackage{multirow}
\usepackage{booktabs}
\usepackage[mathlines,displaymath]{lineno}
\usepackage{appendix}
\journal{Journal of Banking and Finance}

\begin{document}
\onehalfspacing
\begin{frontmatter}
\title{}
\author[inst1]{Yong Bian}
\ead{yongb@mail.zjgsu.edu.cn}

\author[inst2]{Xiqian Wang}
\affiliation[inst2]{organization={School of Economics and Management, Beijing University of Technology}}
\ead{wangxiqian@bjut.edu.cn}

\author[inst1]{Qin Zhang\corref{cor1}}
\affiliation[inst1]{organization={Tailong Finance School, Zhejiang Gongshang University},
            addressline={18 Xuezheng Street, Qian Tang District}, 
            city={Hangzhou},
            postcode={310018}, 
            state={Zhejiang},
            country={China}}
\ead{zhangqinwenya@mail.zjgsu.edu.cn}
\cortext[cor1]{Corresponding author}

\title{How Does China's Household Portfolio Selection Vary with Financial Inclusion?}
\begin{abstract}
Portfolio underdiversification is one of the most costly losses accumulated over a household's life cycle. We provide new evidence on the impact of financial inclusion services on households' portfolio choice and investment efficiency using 2015, 2017, and 2019 survey data for Chinese households. We hypothesize that higher financial inclusion penetration encourages households to participate in the financial market, leading to better portfolio diversification and investment efficiency. The results of the baseline model are consistent with our proposed hypothesis that higher accessibility to financial inclusion encourages households to invest in risky assets and increases investment efficiency. We further estimate a dynamic double machine learning model to quantitatively investigate the non-linear causal effects and track the dynamic change of those effects over time. We observe that the marginal effect increases over time, and those effects are more pronounced among low-asset, less-educated households and those located in non-rural areas, except for investment efficiency for high-asset households.
\end{abstract}

\vspace{1cm}
\begin{keyword}
Financial Inclusion \sep Investment Efficiency \sep Dynamic Double Machine Learning \sep Asset Allocation \sep Financial Market Participation

\vspace{0.6cm}
\JEL G11 \sep G50 \sep G51 \sep D14
\end{keyword}
\end{frontmatter}

\newpage
\doublespacing
\section{Introduction}
Finance theory predicts that rational households are expected to allocate their risky assets in a well-diversified portfolio \citep{markowits1952portfolio,rubinstein2002markowitz}. Understanding the extent to which households conform to this prediction is crucial for regulating financial products and promoting financial well-being. However, household financial decisions are complex, interdependent, and heterogeneous. Despite substantial evidence of the enduring benefits of allocating investments in riskier assets\footnote{such as stocks, bonds, and bills, which have achieved historical returns in relation to inflation and each other, many investors remain hesitant to take on riskier investments \citep{jorda2019rate,siegel2021stocks}.}, households' investment portfolios do not appear to coincide with the predictions of finance theory as the literature often suggests households tend to underinvest and under-diversify in financial markets \citep{mehra1985equity,campbell2006household,gomes2021household}. In light of the far-reaching consequences of impairments to underinvestment, the question of what drives households' investment decisions has received increased attention among researchers and policymakers alike. Recent research documents that household risk-taking behavior and sub-optimal investment decision is associated with macroeconomic factors, e.g. macroeconomic fluctuation and business circle \citep{knupfer2017formative,luetticke2021transmission}, the social environment, e.g. the influence of peers \citep{hong2004social,kaustia2012peer,bailey2018economic}, cultural influences \citep{haliassos2017incompatible}, latent genetic factors \citep{linnainmaa2011some,calvet2014twin}, income risk and borrowing constraint \citep{guiso1996income,gomes2020portfolio}, personal characteristic e.g. education, age, financial literacy and cognitive abilities, saving goals, gender, age and liability level \citep{campbell2006household,guiso2013household,gomes2021household}. 

In developing countries such as China, a large proportion of the population, especially households in rural areas, is suffering from having difficulty obtaining a wide range of investment opportunities \citep{li2016china,chen2023can}. In addition, the issue of insufficient diversification in household portfolio choice is even more exacerbated as most Chinese individual investors are exposed to extreme portfolio risk (either too little or too much risk) \citep{lu2021digital}, resulting in incompatible with household-consumption smoothing and income inequalities \citep{cocco2005consumption,korniotis2011behavioral}. In recognizing the consequences of under-investment, China's policymakers have pledged to boost credit constraints and other funding measures to individual and private companies who were previously under-served by large state-owned banks to shore up confidence and support the recovery of investment in the financial markets, yet the effectiveness of this policy is a novel subject that deserves thorough research.\footnote{Sources: The State Council, The People's Republic of China, \url{https://english.www.gov.cn/premier/news/2016/07/27/content_281475402898158.htm}}

Against this background, our paper aims to study what drives household investment behavior in China. We propose a hypothesis that having better financial access for households can solve the puzzle of under-investment and improve investment efficiency. To test this hypothesis, we consider using the development of financial inclusion services to proxy relaxation of financial accessibility and assessing the extent to which the variety and intensity of an individual’s financial inclusion affect households' willingness to participate in the financial market, their asset allocation in the portfolio, and investment efficiency. We argue that better financial inclusion penetration makes financial services more accessible for households that were previously underserved by breaking down the traditional barriers and increasing their willingness to participate in the financial markets. Indeed, financial inclusion is attracting considerable interest in both the academic community and government authorities, as it is an important determinant of economic growth and poverty reduction \citep{demirguc2018global}. Households that are financially included are able to increase savings, launch businesses, reach better investment efficiency, and enhance financial well-being \citep{beck2007reaching,dupas2013don,bruhn2014real,angelucci2015microcredit,banerjee2015miracle}.

A number of issues have arisen regarding c First, in contrast to the existing literature where scholars use the regional financial inclusion index to account for individuals' financial inclusion (i.e. Peking University Digital Financial Inclusion Index of China in \cite{huang2022digital} and \cite{lu2023digital}), we argue that simply assigning regional-level data to represent household characteristics is irrational; instead, our view is that the measurement of financial inclusion for the individual household should be more precise to reflect their product awareness, active choice of products, and knowledge of alternatives to formal financial services. We employ China Household Finance Survey (CHFS) data to construct a detailed measure of household individual financial inclusion and relate the measure of financial inclusion of individual households to digital (or online) financial products and services. We pin down four specific measures of financial inclusion \textendash having access to credit cards, third-party digital payments, insurance services, and bank accounts.

Second, although the current literature has largely accounted for the possible bidirectional relationship between financial inclusion and household portfolio choice, estimating the \textit{causal} effects of having access to financial products and services on household investment choice can be biased due to several challenges, such as omitted variables bias, measurement error, non-linearity, and the difficulty of separately identifying the different treatments. This issue is even more exacerbated when one must select the most vulnerable confounders from a large panel of pooled variables \citep{wing2018designing}. Traditional econometric models for causal analysis are confined to examining the linear causal relationship in a low-dimensional parameter environment, raising concerns about the validity of causal inference \citep{athey2017state}. In addition, these studies often overlook aspects of time dynamics, which can lead to misguided policy implications if policy treatments are given in continuous time. We overcome the challenges described above by employing dynamic double machine learning (DDML), which is an extension of the Double Machine Learning framework of \cite{ccddhn2017} that combines machine learning techniques with econometric methods to address causal inference problems. DDML is particularly useful for handling time-varying effects, providing robustness to confounding factors, accounting for high-dimensional data and adapting to nonlinearity \citep{lewis2020double}.

We \textit{hypothesize} that (i) having access to financial inclusion services encourages households to participate in the financial market; (ii) having access to financial inclusion services increases the diversification of household equity investments; (iii) having access to financial inclusion services increases the \textit{efficiency} of diversification of household equity investments; and (iv) the effects of financial inclusion services on household portfolio diversification increase over time. We merge data from the China Household Finance Survey (CHFS) of \cite{gan2014data} to fully reflect household characteristics and the basic situation of household finance, including gender, demographic background, assets and investments, income and expenditure, insurance, and much other micro-level information. Our detailed data allows us to causally analyze if, and to what extent, the effect of better financial inclusion on household portfolio asset allocation and how these effects vary among different subgroups of individuals. 

To preview our main results, our baseline fixed effect model provides strong evidence that having access to financial inclusion increases the likelihood that households engage in the financial markets, allocate their investment to risky assets and improve investment efficiency, after controlling for various household head and household characteristics. The effects are statistically significant and of sizable magnitude. Considering there might exist possible non-linear and interactional effects that can not be captured by our baseline model, we employ double machine learning (DML) of \cite{dml2018}. The results of DML support the notion that there are positive impacts of having better financial access on household risk-taking investment behavior and investment efficiency. We further strengthen our findings through dynamic analysis of the effects over a three-year panel. We show that, based on 2015 to 2019 data, there are consistent upward trajectories in household risky asset purchase and investment efficiency from the provision of financial inclusion, suggesting that gradual penetration of financial inclusion services is necessary to promote household financial well-being. Further analyses indicate our results are robust to the alternative measure of the financial inclusion index and using city average financial inclusion as an instrument. In addition, we show results are heterogeneous by households' residency, educational background and asset level.

This study contributes to the literature in several ways. First, we build on a burgeoning literature that investigates what drives household financing behavior in China, especially filling the research gap on how financial inclusion and its accessibility help to solve the low-participation rate in the stock market puzzle. The most important fact that emerges from the analysis of the asset allocations of households is that a large fraction of the population simply does not own any risky asset, especially equity \citep{badarinza2016international,gomes2021household}. For example, research shows that, based on data from the Eurosystem Household Finance and Consumption Survey (HFCS), national household surveys from Australia, Canada, and the U.S., equity market participation rates were below 50\% in all countries except for Sweden \citep{badarinza2016international}. Limited stock market participation comes from various reasons such as ambiguity and risk preference (for instance a lack of trust in corporate fraud), the influence of peers, high transaction costs, and insufficient financial literacy \citep{choi2020matters,peijnenburg2018life,pagel2018news,gomes2005optimal,guiso2008trusting,hong2004social}, income uncertainty, and borrowing constraints \citep{bonaparte2014income,calvet2014twin}.\footnote{For instance, \cite{changwony2015social} and \cite{kanagaretnam2022trusting} find that there has been a persistent lack of direct participation in the stock market, despite the increasing popularity and efficiency of indirect investment opportunities through vehicles such as mutual funds. This could be because there is a lack of trust in corporate fraud, damaging general trust in the stock market. In addition, households who bear more liability are more likely to allocate their money to safe investments \citep{gomes2021household}.} Our paper adopts a perspective that examines how the promotion of inclusive finance can incentivize households to engage with the financial market. Our definition of the financial market is more comprehensive than what is typically found in existing literature. We encompass a broad spectrum of financial assets, extending beyond traditional stocks, as highlighted in previous studies such as \cite{changwony2015social} and \cite{liang2015social}.

Second, compared with existing literature, our measure of households' level of financial inclusion is more precise. For example, based on a panel of nationwide data from the CHFS and Peking University Digital Financial Inclusion Index, \cite{lu2021digital} and \cite{lu2023digital} find that digital financial inclusion significantly increases the diversification of stock investments and decreases the probability of households taking extreme portfolio risks. Our study differentiates from previous studies in the sense that the measures of financial inclusion are drawn from household surveys, which are more intuitive in representing the status of inclusive finance for individual households than assigning a common index for individuals who live in the same region.

Finally, our results support the usefulness of machine learning algorithms in casual analysis, especially when one needs to account for non-linear, interactional, and dynamic effects. Traditional econometrics methods typically rely on a precise selection of control variables. If the effect might be driven by some unobserved factors or some common time trend in the variations of confounders, the exclusion of those unobserved factors leads to inconsistent estimated coefficients and biased policy implications. Scholars find machine learning algorithms provide robustness in estimating average causal effects. For instance, \cite{yang2020double} examine double machine learning in conjunction with gradient boosting within the context of its substantial impact on audit quality. Their proposed DML framework introduces regularization bias while accommodating flexible functional forms, consistently estimating average causal effects within the realm of high-dimensional covariate settings. Their findings exhibit robustness concerning the hyperparameters utilized in gradient boosting, a marked contrast to the outcomes yielded by propensity score matching. In a similar vein, \cite{knaus2022double} leverage DML to assess the effectiveness of the Swiss Active Labor Market Policy. Their investigation reveal the resilience of DML when compared to alternative causal methodologies, including semi-parametric mediation analysis. This framework serves as an exemplar of the potential applications of DML-based approaches for evaluating programs under the assumption of unconfoundedness. It accommodates complex functions of confounding variables and offers a blueprint for further analyses in similar domains.

The rest of the paper is organized as follows. Section \ref{data} describes our data used in empirical analysis. Section \ref{Result} presents our identification approach and discusses the main empirical results. Section \ref{heterogeneity} provides results concerning robustness and individual heterogeneity. Section \ref{Conclusion} concludes. 

\section{Data}\label{data}
We collect data from the China Household Finance Survey (CHFS), which was initially conducted by the Southwest University of Finance and Economics in 2011 and has been disclosed every two years since then. CHFS has been well-recognized in the research community as a reliable dataset and has been used extensively in top journals such as \cite{chen2019social} (\textit{Review of Finance}), \cite{feng2019financial}, \cite{li2021entrepreneurship} (\textit{Journal of Empirical Finance}), and \cite{yang2023digital} (\textit{Journal of Banking and Finance}). The survey is designed to be representative of household finance by employing scientific sampling, modern survey techniques such as computer-assisted personal interviewing (CAPI), and survey management methods to collect demographic characteristics, assets and liabilities, income and consumption, insurance and protection, and other aspects of micro-level information on households. It contains detailed inquiries about various financial assets such as bank deposits, stocks, bonds, funds, and non-financial assets such as housing and vehicles, providing good data support for this article's research. 

In accordance with our fourth hypothesis that the effects of financial inclusion are dynamic over time, we employ the 2015, 2017, and 2019 waves of the survey and keep track of the households who are consistently presented in all three years. In addition, we exclude observations in the 1\% tails for key variables such as the Sharpe ratio, resulting in a balanced panel dataset of 5,694 representative households.

\subsection{Measuring Financial Market Participation, the Choice of Household Assets and Its Efficiency}
We consider using three variables to study how Chinese households' investment decisions are affected by having access to financial inclusion. For the first dependent variable, we examine the effects of financial inclusion on whether households actively participate in the financial market by creating a binary variable financial market participation (FMP), denoting whether households possess a positive amount of risky financial assets (=1 if yes; otherwise =0). Unlike prior literature that mainly focuses on stock market participation (e.g. \citep{changwony2015social,liang2015social}), we consider a broader range of financial markets, including the stock market, fixed-income securities, bank financial products, insurance, trusts, and derivatives. Therefore, our measures of risky assets and financial markets are more comprehensive. More specifically, risky assets can be categorized as follows: (1) Risky assets in formal markets: This category encompasses assets such as stocks, funds, fixed-income securities, financial derivatives, financial wealth management products, foreign exchange, and gold. (2) Risky assets in informal financial markets: This category primarily involves private lending activities, including borrowing from family members, friends, business partners, and shadow credit providers. Meanwhile, risk-free assets encompass holdings like cash, cash in stock accounts, government bonds, demand deposits, and term deposits.

To examine how household risk preferences in the financial markets are influenced by financial inclusion penetration, we measure the willingness of households to allocate investment between risky and risk-free assets \textendash the risky asset ratio, which is simply the proportion of risky assets in relation to total assets. The definitions of risky assets are consistent with financial market participation.

The third dependent variable measures the impacts of having access to financial inclusion on the efficiency of household asset allocation and investment decisions. Following \cite{gaudecker2015does} and \cite{wu2021digital}, we opt for the self-constructed Sharpe ratio as the measure of the efficiency of household portfolio choice. To construct the Sharpe ratio for each household, we restrict the household's portfolio to only risk-free assets and risky assets. With the risky assets class, we further divide them into two sub-categories according to the level of return and risk: bond-based and stock-based assets. Bond-based assets include bonds, financial management products, Internet financial management products, bond-oriented funds, and currency-oriented funds, whereas stock-based assets include stocks, equity-oriented funds, derivatives, non-RMB assets, precious metals, and other higher-risk financial products.

We calculate the Sharpe ratio of the household financial asset portfolio as:
\begin{equation}
\begin{split}
\textrm{Sharp\_ratio}_{i}=\frac{E(R_{pi})-R_{f}}{\delta_{pi}} \\
E(Rp_{i})=\sum_{j=1}^{m}w_{j}R_{j} \\
\delta_{pi}=\sqrt{\delta_{pi}^2}=\sqrt{\sum_{j=1}^{N} \sum_{k=1}^{N} w_{j}w_{k} \sigma(R_{j}, R_{k})}
\end{split}
\label{sharp_ratio}
\end{equation}
in Equation \ref{sharp_ratio}, $E(R_{pi})$ and $\delta_{pi}$ represent the expected return and standard deviation of the households' financial portfolio, respectively. $R_f$ represents the risk-free rate of return, measured by the interest rate of a one-year term deposit. $w_i$ represents the proportion of each type of financial asset in the total financial assets of the household, and $N$ represents the number of types of financial assets invested by the household. $\sigma(R_{j}, R_{k})$ represents the covariance between the returns of various assets. If $j=k$, it represents the variance of that type of financial asset. Since the Chinese household financial survey data only provides information on the amount of household asset allocation and no return data are available, we need to approximate the rate of return and risk of each asset. The annual yield of deposit-based assets is replaced by the benchmark interest rate of one-year term deposits published by China's Bank of People and they are risk-free; the annual yield and risk of bond-based assets are set to the annual yield and standard deviation of the CSI Comprehensive Bond Index; the annual yield and risk of stock-based assets are set to the annual yield and standard deviation of the SSE Composite Index and the SZSE Component Index weighted by trading volume. All calculations are based on data from January 2003 to the year of the survey was conducted, which are collected from the CSMAR database. 

\subsection{Measuring Financial Inclusion}
To construct a measure of household-level financial inclusion, we draw on the following questions related to four types of financial inclusion services in CHFS:

\leftskip=1cm
\rightskip=1cm
``Have you or your families had access to financial services related to financial inclusion for the credit card(s) (or third-party digital payment(s), or bank account(s), or commercial insurance)?"\footnote{Note that CHFS survey teams ask individual questions related to each financial service.}

\leftskip=0cm
\rightskip=0cm
\noindent For each of the four questions, we assign a value of 1 if the household answers ``Yes, I and (or) my families have access to this financial service" and a value of 0 if the household answers ``No''. A comprehensive financial indicator is derived by the simple average of the four individual financial inclusion measures.

\subsection{Control Variables}
As the choice of household risky asset allocation and investment efficiency can be related to many factors that may also influence financial inclusion, we control for two characteristics: (1) household head's characteristics, covering age, gender, marital status, whether engaged in business and health status; (2) household's characteristics, including the number of elderly and children in the family, as well as the total population of the household. In addition, considering that the opening conditions for most bank accounts and applying for credit cards require individuals to be fully capable of civil conduct and must be aged over 18 years, we delete samples under 18 years old. Finally, we exclude invalid samples with missing observations and retain individuals who appear in all years, resulting in a balanced panel of a total of 17,082 valid samples.

Table \ref{variable_definition} provides variable descriptions for the characteristic control variables. Table \ref{variable_statistics} shows summary statistics for all the variables.

\begin{table}[h!]
\centering
\def\arraystretch{1.4}
\caption{Variables Definition}
\scalebox{0.85}{
\begin{tabular}{ll}
\toprule
Variable Name & \multicolumn{1}{c}{Definition} \\
\midrule
age & Age of the Household Head(the Head): Year of Survey minus Year of Birth\\
male & Head being Male: 1=Male; 0=Female \\
marriage & \makecell[l]{Marital Status of the Head: 1=Married, Cohabiting\\ 0=Unmarried, Separated, Divorced, Widowed} \\
ind\_commer & Engaged in Industry and Commerce: 1=Yes; 0=No \\
edu & \makecell[l]{Education Level of the Head: None=1, Primary School=2,\\ Junior High=3, High School=4,Secondary Vocational School=5, \\Junior College (with Associate's Degree)=6, \\Bachelor's Degree=7, Master's Degree=8, Doctorate's Degree=9} \\
health & Health Status of the Head: 1=Very Good; 2=Good; 3=Normal; 4=Bad; 5=Very Bad\\
oldsum & Number of Households Aged 65 and Above \\
youngsum & Number of Households Aged 14 and Below \\
family\_size & Number of Members in the family\\
asset & \makecell[l]{High-asset Group\textendash Asset Greater than 3 million Yuan\\Middle-asset Group \textendash Asset Between Half Million to 3 Million\\Low-asset Group\textendash Asset Less Than Half Million Yuan} \\
rural & 1=Living in Rural Region; 0=Living in Urban Region \\
east & 1=Resident in Eastern Region, 0=Otherwise\\
middle &  1=Resident in Middle Region, 0=Otherwise \\
west & 1=Resident in Western Region, 0=Otherwise \\
risk\_prefer & 1=Prefer High Risk \& High Return ; 0=Prefer Low Risk \& Low Return Investment\\
fina\_know & Financial Literacy: 1=Correctly Answer Two Financial Questions; 0=Otherwise \\
\bottomrule
\end{tabular}}
\parbox{0.9\textwidth}{\footnotesize%
\vspace{2eX} 
Note: Two Financial Questions: 1. Assuming the bank's annual interest rate is 4\%, if you deposit 100 yuan in a one-year fixed-term account, the principal and interest you'll receive after 1 year will be? 2. Assuming the bank's annual interest rate is 5\%, and the annual inflation rate is 3\%, how will the purchasing power of 100 yuan deposited in the bank change after one year?} 
\label{variable_definition}
\end{table}

\begin{table}[h!]
\centering
\def\arraystretch{1.3}
\caption{Summary Statistics}
\begin{tabular}{lccccc}
\toprule
Variable &Median & Mean & Standard Deviation & Min & Max \\
\midrule
FMP &1 & 0.581 &0.493 &0&1\\
Risky Asset Ratio & 0.019 & 0.201 & 0.287 & 0 & 1 \\
Sharpe Ratio & 0.395 & 0.438 & 0.156 & 0.364 & 3.905 \\
credit card  & 0  &0.150 &0.357 &0 & 1\\
digital payment &0  &0.205 &0.403 &0 & 1\\
bank account &1 & 0.891 &0.311&0 & 1\\
commercial insurance & 0 & 0.090 &0.286&0&1\\
FA index & 0.250 & 0.334 & 0.204 & 0 & 1 \\
age & 56 & 56.830 & 12.696 & 19 & 100 \\
male & 1 & 0.798 & 0.401 & 0 & 1 \\
marriage & 1 & 0.887 & 0.317 & 0 & 1 \\
ind\_commer & 0 & 0.135 & 0.341 & 0 & 1\\
edu & 3 & 3.301 & 1.474 & 1 & 9 \\
health & 3 &2.687 &0.972 & 1 & 5\\
oldsum & 0 &0.660 &0.835 &0 & 4\\
youngsum &0 & 0.403 & 0.706 &0 & 7\\
family\_size &3 & 3.346 &1.559 &1 &19\\
low\_asset &1&0.591&0.492 & 0 & 1\\
middle\_asset & 0& 0.266 &0.442&0 & 1\\
high\_asset & 0 & 0.144 & 0.351& 0 & 1\\
rural & 0 & 0.396 & 0.489 & 0 & 1\\
east & 0 & 0.460 &0.498 & 0 & 1\\
middle & 0 & 0.312 &0.463 & 0 & 1\\
west &0 & 0.228 &0.419 &0 & 1\\
risk\_prefer & 0 &0.076 &0.265 &0&1\\
fina\_know &0 & 0.099 &0.298 &0 &1\\
\bottomrule
\end{tabular}
\parbox{0.9\textwidth}{\footnotesize%
\vspace{2eX} 
Note: FA index is the comprehensive financial inclusion indicator, calculated by the average of the four individual financial inclusion measures: (whether have access to) credit card(s), digital payment(s), bank account(s) and commercial insurance.} 
\label{variable_statistics}
\end{table}

\section{Empirical Results}\label{Result}
\subsection{Effects of Financial Inclusion on Household Asset Allocation and Investment Efficiency}
We start by investigating the effect of household-level financial inclusion on households' decision to allocate their fraction of investment into risk-free and risky assets as well as investment efficiency. Our hypothesis is that, with the promotion of financial inclusion to ease borrowing constraints, households can be more inclined to participate in the financial market and allocate their investment in risky assets with potentially higher returns as they might perceive financial inclusion (such as insurance products) as a sense of security to hedge risk, making them more likely to take on riskier financial endeavors. As a result, a better asset allocation strategy allows them to achieve desired financial goals while managing risk effectively, yielding improved investment efficiency \citep{yang2022fintech}. 

To examine whether the promotion of financial inclusion encourages households to participate in the financial market, allocate more investment into risky assets, and yield better investment efficiency, we consider employing the fixed effects model by regressing the binary variable of whether households participate in the financial market, the ratio of risky assets to total assets for each individual, and Sharpe ratio on the measure of our self-constructed financial inclusion index in a panel data regression setting as follows:
\begin{align}
Y_{i, j, t}=\alpha+\beta_1 \cdot \mathrm{FA} \mathrm{index}_{i, j, t}+\sum_{i, j, t} \gamma_{i, j, t} \cdot \operatorname{Control}_{i, j, t}+\varphi_t+\varphi_j+\epsilon_{i, j, t} \label{Fixed-regression}
\end{align}
where $Y_i$ are the outcome variables of financial market participation, risky asset ratio, and Sharpe ratio. $i$, $j$, and $t$ refer to household, province, and year, respectively. Considering that there might exist factors that influence the connection between financial inclusion and households' finance behaviors, we account for various household head's and household characteristics, such as age, gender, marital status, the number of elderly and children in the family, as well as the total population of the household. In addition, time-fixed effects and province-fixed effects are indicated as $\varphi_t$ and $\varphi_j$. $\epsilon_{i, j, t}$ is model residual.

Table \ref{FEresult} presents the baseline results illustrating the influence of financial inclusion on household financial market participation, financial asset allocation, and investment efficiency. Observed in columns (1), (3), and (5) of Table \ref{FEresult}, we find financial inclusion penetration positively and significantly correlates with households' willingness to participate in the financial markets, risky asset fraction, and investment efficiency, which is consistent with our proposed hypothesis. The correlation is of a sizable magnitude. In particular, representative households switching from having none of any four financial inclusion services (digital payment(s), credit card(s), bank account(s), and commercial insurance) to gaining access to all of them leads to increases in financial market participation by 83.7\% (\textit{t-stat}=48.2), the risky asset ratio by 32.8\% (\textit{t-stat}=32.8) and Sharpe ratio by 0.167 (\textit{t-stat}=27.8). 

When we control for characteristics and other demographic factors, the estimated coefficients from the fixed effects model in columns (2), (4), and (6) remain positive and significant, further supporting our hypothesis. The estimates on the probability of financial market participation, risky asset ratio, and Sharpe ratio reduce to 47\% (\textit{t-stat}=23.5), 19\% (\textit{t-stat}=15.83) and 0.125 (\textit{t-stat}=20.83), respectively. Given that the average probability of households engaging in the financial markets is 58.1\%, the average portion of risky assets a household purchased is 20.1\%, and the average Sharpe ratio is 0.438, such improvements are sizable. Overall, these results strongly support that the penetration of financial inclusion plays a critical role in driving households' investment decisions and promoting investment efficiency, which in turn, no doubt will lead to increased financial well-being for both households accumulated over the household life-cycle as well as society as a whole. 

\begin{table}[h!]
\centering 
    \caption{Fixed Effects: How Financial Inclusion Influences Households' Investment Behavior}\label{FEresult}
\scalebox{0.7}{
\begin{tabular}{lccccll}
  \toprule
 &\multicolumn{2}{c}{FMP}&\multicolumn{2}{c}{Risky Asset Ratio}&  \multicolumn{2}{c}{Sharpe Ratio}\\
    \cmidrule(lr{.75em}){2-3} \cmidrule(lr{.75em}){4-5}\cmidrule(lr{.75em}){6-7}
     & (1) & (2) & (3) & (4) &   (5) &(6)\\ 
  \cmidrule{2-7}
(Intercept) & 0.302***&  0.139***&0.091***& 0.043* &  0.382*** &0.333*** \\
                      &(0.007)&(0.036)& (0.004)&(0.023)&  (0.002)&(0.012)\\
  FA\_index & 0.837***& 0.470***& 0.328*** &0.190***&  0.167***&0.125*** \\
                   &(0.017)& (0.020)& (0.010)  &(0.012)&  (0.006) &(0.006)\\
age               &&    -0.002***&&0.000***&  &0.000***   \\
                    &&    (0.000)&&(0.000)&  &(0.000)   \\
male             &&     0.008&&0.000&  &0.001    \\
                  &&     (0.009)&&(0.006)&  &(0.003)    \\
marriage        &&       0.014&&-0.006&  &-0.003     \\
                  &&      (0.011)&&(0.007)&  &(0.004)   \\
ind\_commer\       &&      0.044***&&0.013**&  &-0.006* \\  
                  &&      (0.010)&&(0.006)&  &(0.003)   \\
edu                &&      0.034***&&0.016***&  &0.015***    \\
                     &&     (0.003)&&(0.002)&  &(0.001)   \\
health              &&       -0.008**&&-0.002&  &0.003***   \\
                 &&       (0.004)&&(0.002)&  &(0.001)    \\
oldsum            &&        -0.002&&0.003&  &0.006***    \\
                   &&      (0.005)&&(0.003)&  &(0.002)    \\
youngsum           &&        -0.025***&&0.000&  &0.003   \\
                     &&      (0.006)&&(0.004)&  &(0.002)   \\
family\_size           &&     0.021***&&0.002&  &-0.004***  \\
                      &&      (0.003)&&(0.002)&  &(0.001)\\
high\_asset          &&      -0.030&&0.028***&  &0.035***   \\
                           &&(0.011)&&(0.007)&  &(0.003)   \\
middle\_asset          &&    0.067***&&0.054***&  &0.017***   \\
                        &&     (0.009)&&(0.005)&  &(0.003)    \\
fina\_know              &&   0.021*&&0.025***&  &0.025***  \\
                       &&      (0.012)&&(0.007)&  &(0.004)    \\
risk\_prefer            &&    0.036***&&0.047***&  &0.066***   \\
                            &&   (0.013)&&(0.008)&  &(0.004)   \\
east                       &&  0.028&&0.047***&  &0.047***   \\
                         &&   (0.027)&&(0.017)&  &(0.009)    \\
middle                     &&     0.077***&&0.029*&  &0.008    \\
                            &&   (0.025)&&(0.016)&  &(0.008)    \\
rural                       && -0.028***&&-0.030***&  &-0.011***\\
                            && (0.008)&&(0.005)&  &(0.003)\\
year2017                   && 0.208***&&0.034*** &  &-0.024***\\
                           && (0.009)&&(0.006)&  &(0.003)\\
year2019                  && 0.340***&&0.038***&  &-0.063***\\
                          && (0.009)&&(0.006)&  &(0.003)\\
                              &&&&&  &\\
Obs. & 17,082& 17,082& 17,082& 17,082&  17,082 &17,082\\ 
Province FE & No&Yes&No&Yes&  No&Yes\\
\bottomrule
\end{tabular}}
\parbox{0.9\textwidth}{\footnotesize%
\vspace{2eX} Note: Reported are estimated coefficients of the effects of financial inclusion on households' investment behavior using fixed effect OLS. Asterisks indicate ***0.1\%, **1\%, and *5\% significance levels for the $p$-value. Standard errors are reported in parentheses. Financial market participation is assigned a value of 1 if the household owns any risky assets and a value of 0 otherwise. FMP is estimated using a linear probability model rather than a logit or probit model. The rationale is that we limit our dependent variable financial inclusion index to have a range from 0 to 1, with 0 being having none of financial access and 1 being having all four financial access; therefore, the economic interpretation of the FMP estimate can be narrowed down.} 
\end{table}

\newpage
\subsection{Are Effects Non-linear?}
After documenting positive relationships among financial inclusion, households' asset allocation choice, and investment efficiency, we are particularly interested in asking whether there exist possible non-linear and interactional effects. This is motivated by the consideration that the effects might still be driven by some unobserved factors at the province level, interaction terms or some common time trend in the variations of control variables affecting households' asset allocation decisions, investment efficiency, and the promotion of financial inclusion \citep{hong2020fintech}. If we left those unobserved factors behind in Equation \eqref{Fixed-regression}, the estimated effects are inconsistent, leading to biased policy implications. However, if we include all possible interactional effects, solving Equation \eqref{Fixed-regression} would be infeasible. We address the challenges posed by a large number of variables and their possible functional forms in Equation \eqref{Fixed-regression} by leveraging Double Machine Learning of \cite{dml2018} while retaining the rigor of traditional econometric methods. It offers benefits such as dimensionality reduction, robustness to outliers, improved model selection, and handling of non-linear relationships, making it a powerful tool for analyzing high-dimensional data compared to traditional OLS estimation. It reads as:

\begin{subequations}
\begin{align}
& Y_{i}=\theta_{0} \operatorname{FAindex}_{i, j, t}+g_{0}\left(\operatorname{Control}_{i, j, t}\right)+U_{i}, E\left[U_{i} \mid X_{i}, \operatorname{FAindex}_{i, j, t}\right]=0 \label{eq1} \\
& \operatorname{FAindex}_{i, j, t}=m_{0}\left(X_{i}\right)+V_{i}, E\left[V_{i} \mid X_{i}\right]=0 \label{eq2}
\end{align}
\end{subequations}
where $Y_i$ are the outcome variables financial market participation, risky asset ratio and Sharpe ratio, $U_i$ and $V_i$ are model errors from \eqref{eq1} and \eqref{eq2}, respectively. Equation \eqref{eq1} is a semi-parametric (or partial linear) Conditional Expectation Function (CEF), with $\operatorname{FAindex}_{i, j, t}$ linearly added and all the other control variables lump summed in a nonparametric function $\mathrm{g}_{0}(*)$. Control variables are allowed to be correlated with treatment variables as denoted by Equation \eqref{eq2}, another CEF. The nuisance parameters in $g_{0}(*)$ or $m_{0}(*)$ are not of interest, but the precision of $g_{0}(*)$ and $m_{0}(*)$ are critical, as they can be very well estimated by nonparametric machine learning methods. 

DML builds a Neyman Orthogonal score as\footnote{Introduced by Neyman (1959)}: 
\begin{align}
\psi\left(Y_{i}, \operatorname{FAindex}_{i, j, t}, \operatorname{Control}_{i, j, t}; \theta, \eta\right), \,\,\,\,\eta=(g, m)
\end{align}
which satisfies a first-order moment condition of $E(\psi(Y_{i}, \operatorname{FAindex}_{i, j, t}, \operatorname{Control}_{i, j, t}; \theta_{0}, \eta_{0}))=0$ and an orthogonal condition of $\partial_{\eta} E(\psi \operatorname{FAindex}_{i, j, t}, \operatorname{Control}_{i, j, t}; \theta_{0}, \eta_{0}))\left[\eta-\eta_{0}\right]=0$ conditional on $\eta_{0}=(g_{0}, m_{0})$. 

DML estimator differs from traditional econometric models by a sample splitting procedure which allows the part of data used in estimating $g_{0}(*)$ and $m_{0}(*)$ to be different from those used to obtain the final estimate, resulting in reduced the overfitting bias. $\tilde{\theta}_{0}$ is obtained from the moment condition of the orthogonal score function:
\begin{align}
\frac{1}{K_{k=1}^{K}} \sum_{n k}\left[\psi\left(Y_{i}, \operatorname{FAindex}_{i, j, t}, \operatorname{Control}_{i, j, t}; \tilde{\theta}_{0}, \hat{\eta}_{O k}\right)\right]=0.
\end{align}

\begin{table}[h!]
    \centering 
    \caption{DML: How Financial Inclusion Influences Households' Investment Behavior}\label{DMLresult}
    \def\arraystretch{1.35}
    \scalebox{0.8}{
    \begin{tabular}{lccccc}
    \toprule
 Panel A &\multicolumn{5}{c}{FMP}\\
\midrule
     & (1) & (2) & (3)&(4) & (5)\\
      \cline{2-6}
    15-15 &0.498***&0.391***&0.170***&-0.039*&0.197***\\
          &(0.043)&(0.024)&(0.020)&(0.024)&(0.022)\\
    17-17 &0.373***&0.193***&0.071***&0.108***&0.130***\\
          &(0.036)&(0.019)&(0.017)&(0.025)&(0.019)\\
    19-19 &0.485***&0.322***&0.059***&0.112***&0.073***\\
          &(0.023)&(0.012)&(0.011)&(0.016)&(0.011)\\
        \bottomrule
 Panel B& \multicolumn{5}{c}{Risky Asset Ratio}\\
\midrule
& (1) & (2) & (3)&(4) & (5)\\
      \cline{2-6}
     15-15 & 0.263***& 0.163***& 0.076***& -0.001& 0.111***\\
 & (0.026)& (0.021)& (0.012)& (0.012)& (0.014)\\
 17-17 & 0.198***& 0.086***& 0.037***& 0.039***& 0.105***\\
 & (0.026)& (0.017)& (0.012)& (0.013)& (0.015)\\
 19-19 & 0.200***& 0.114***& 0.066***& 0.016& 0.063***\\
 & (0.020)& (0.010)& (0.012)& (0.011)& (0.014)\\
 \bottomrule
  Panel C & \multicolumn{5}{c}{Sharpe Ratio}\\
\midrule
& (1) & (2) & (3)&(4) & (5)\\
      \cline{2-6}
    15-15 &0.108***&0.057***&0.040***&0.014***&0.029***\\
          &(0.016)&(0.017)&(0.008)&(0.004)&(0.008)\\
    17-17 &0.136***&0.104***&0.031***&0.017***&0.032***\\
          &(0.017)&(0.014)&(0.008)&(0.002)&(0.010)\\
    19-19 &0.115***&0.039***&0.067***&0.014***&0.050***\\
              &(0.014)&(0.005)&(0.011)&(0.003)&(0.012)\\
 \bottomrule
    \end{tabular}}
\parbox{0.9\textwidth}{\footnotesize%
\vspace{2eX} Note: Reported are estimated coefficients of the effects of financial inclusion on households' investment behavior using the DML model. Asterisks indicate ***0.1\%, **1\%, and *5\% significance levels for the $p$-value. Standard errors are reported in the parentheses. Each row corresponds to the effect of financial inclusion of the indicated year. Column (1) displays the impact of the FA index. Column (2) illustrates the influence of whether the household uses digital payment(s). Column (3) demonstrates the impact of household ownership of credit card(s). Column (4) reflects the effect of whether the household possesses bank account(s). Column (5) indicates the influence of whether the household has invested in commercial insurance.}    
\end{table}

In the DML approach, we model data from different years separately while controlling for provincial variations. This strategy addresses the heterogeneity arising from distinct years and provinces. By adequately accounting for essential demographic and characteristic variables, the influence of financial inclusion penetration should be uncorrelated with how households allocate financial assets and their efficiency within each year. This guarantees the consistency of our cross-sectional DML analysis when considering different years independently, providing reliable effect estimates.

Table \ref{DMLresult} presents the results of the DML estimation. The table documents the overall impacts of the financial inclusion index (FA index) in columns (1) and (6), while the analyses for each individual financial inclusion service are delved into subsequent columns (2) to (5) and columns (6) to (10).\footnote{More specifically, columns (2) and (7) elucidate the effect of households accessing third-party digital payment(s); columns (3) and (8) examine the effect of households having credit card(s); columns (4) and (9) explore the effects of households holding bank account(s); while columns (5) and (10) investigate the effect of households purchasing commercial insurance.} The results from DML estimations indicate that, although there are slight variations in the effects of the FA index across the three years on financial market participation, risky asset ratio, and the Sharpe ratio, these estimates are consistent with the findings from the fixed effects models detailed in Table \ref{FEresult}. The FA index leads to average increases in financial market participation by 49.8\% in 2015, 37.3\% in 2017 and 48.5\% in 2019; enhances risky asset ratio by 26.3\% in 2015, 19.8\% in 2017 and 20.0\% in 2019; and elevates the Sharpe ratio by 0.108 in 2015, 0.136 in 2017 and 0.155 in 2019. 

Regarding each individual financial inclusion service for households in columns (2)-(5) and (7)-(10) of Table \ref{DMLresult}, we find conclusive evidence that better financial access to financial services has a positive impact on households' risky-taking investment behavior and investment efficiency across all three years. Compared to owning bank accounts and purchasing commercial insurance, having access to digital payments and possession of credit cards play more pivotal roles in the overall impact of financial inclusion. These findings offer valuable insights for potential policy implications and further discussions. 

\subsection{Time Dynamic of Effects}
While the DML method effectively provides consistent effect estimates under nonparametric and non-linear settings regarding the nuisance parameters associated with controls, it is limited to cross-sectional analyses. However, in some circumstances, scholars' and policy-makers' interests may lie in comprehending the dynamic impact of a policy over time. Financial inclusion, being a multifaceted policy rather than a one-time treatment, necessitates an exploration of its effects over a period. Therefore, taking time dynamic effects into our analysis is a more reliable approach to depict the how real-world manifests. We focus on detecting dynamic treatment effects on terminal outcomes within our recorded timeframe\textendash specifically, how the penetration of financial inclusion over years influences household asset allocation and its efficiency in the year 2019. To evaluate the dynamic effects, we employ the Dynamic Double Machine Learning method proposed by \cite{lewis2020double}.


We consider a dynamic structure:
\begin{align}
\operatorname{Control}_{i, j, t} & =(A+C\operatorname{Control}_{i, j, t-1})\operatorname{FAindex}_{i, j, t-1}+B\operatorname{Control}_{i, j, t-1}+ \epsilon_{i,j,t} \\
\operatorname{FAindex}_{i, j, t} & =\alpha \operatorname{FAindex}_{i, j, t-1}+(1-\alpha)D\operatorname{Control}_{i, j, t}+\zeta_{i,j,t} \\
Y_{i,j,t} & =\left(\sigma^{\prime} \operatorname{Control}_{i, j, t}+1\right) \cdot e \operatorname{FAindex}_{i, j, t}+f\operatorname{Control}_{i, j, t}+\eta_{i,j,t}
\end{align}
If this process is applied recursively on $t=\{1, \ldots, m\}$, then we can write:
$$
Y_{i,j,m}=\psi_t^{\prime}\operatorname{FAindex}_{i, j, t}+\sum_{q=t+1}^m \psi_q^{\prime}\operatorname{FAindex}_{i, j, q}+\mu^{\prime} B^{q-t} \operatorname{Control}_{i, j, t}+\sum_{q=t+1}^m \mu^{\prime} B^{m-q} \eta_q+\epsilon_m
$$
Since the random shocks $\left\{\eta_q\right\}_{q=t+1}^m$ and $\epsilon_m$ are independent of $\operatorname{Control}_{i, j, t}, \operatorname{FAindex}_{i, j, t}$ and have mean zero, thus the following conditional moment restriction is satisfied:
$$
\mathbb{E}\left[Y_{i,j,m}-\psi_t^{\prime} \operatorname{FAindex}_{i, j, t}-\sum_{q=t+1}^m \psi_q^{\prime}\operatorname{FAindex}_{i, j, q}-\mu^{\prime} B^{m-t} \operatorname{Control}_{i, j, t} \mid \operatorname{FAindex}_{i, j, t}, \operatorname{Control}_{i, j, t}\right]=0 .
$$
The marginal effect of certain period financial inclusion on the final term outcomes can be drawn from solving the moment condition.

Table \ref{DDMLresult} presents the results of the Dynamic DML modeling. In Panel A, we document the dynamic effects across all three years, while Panels B and C provide the dynamic effects for two specific years. Columns (1), (6), and (11) depict the dynamic effects of the FA index, whereas columns (2) to (5), columns (7) to (10), and columns (12) to (15) elucidate the separate effects of individual financial inclusion services. 

Our takeaway findings can be summarized as, in all time periods, there are positive and sizeable marginal effects of financial inclusion on households' financial market participation, their choices of risky asset allocations and investment efficiency, which consistently align with previous results from fixed effects and DML models. In addition,  most of these effects are statistically significant at 1\% level.

Turning into detailed dynamic effects over three-year periods, we observe an upward trend in the marginal effects of financial inclusion penetration for each year on the final year's results, as shown in Figure \ref{DDML_plot}. In addition, the growths of these effects over time are markedly significant. For example, in Panel A column (6) of Table \ref{DDMLresult}, the marginal effects of financial inclusion on household risky asset ratio increase from 4.5\% in the year 2015 to 25.3\% in the year 2019, showing more than five times upward growing trend. The marginal effects on the Sharpe ratio follow a similar pattern\textemdash from 0.067 to 0.142 as shown in Panel A column (11). The increase in the coefficients confirms our presumption of a dynamic structure of financial inclusion penetration on households' investment behavior. The dynamics are recursively progressed as the treatment in the final year is allowed to be affected by the previous treatments and controls, thus it is not surprising the final year's effect is the most pronounced. The results across two-year periods in Panel B and C of Table \ref{DDMLresult} also support the positive and increased dynamic effects of financial inclusion on household portfolio diversification. All effects are statistically significant at the 1\% level except for the two-year panel effect of 15-19 on financial market participation. Compared to the other two measures of household financial allocation behaviors, FMP is more likely to be affected on the final year. The implications drawn from these results suggest that households who never include risky assets in their portfolios take a longer time to make this transition. In contrast, the effects of financial inclusion show quicker impacts, with incremental development over the years on risky asset allocation and investment efficiency.

In addition, we note there are disparities between the 19-19 DML results and the 19-19 Dynamic DML results, which stem from distinct underlying model specifications. In DML, observations are treated as conditionally random, and average treatment effects are derived simultaneously. Consequently, we consider using DML as a cross-sectional analysis. While Dynamic DML allows treatments to evolve with the influence from prior states of other control variables and previous treatments. As a result, the effects differ between the two models.

The dynamics of individual financial inclusion services in columns (2) to (5), columns (6) to (10), and columns (12) to (15) of Table \ref{DDMLresult} are not only consistent with the previous fixed-effect and DML models in demonstrating positive effects but also following the similar patterns of overall findings observed in columns (1), (6), and (11). Notably, among all four financial inclusion services we analyzed, the usage of digital payment(s) and credit card(s) contribute more to overall marginal effects than having bank account(s) and purchasing commercial insurance do.

Overall, our results from Table \ref{DDMLresult} confirm the hypothesis that the effects of financial inclusion on household portfolio diversification increase over time. Considering the importance of household portfolio diversification on their financial wealth and consumption, especially accumulated over a lifetime cycle, our dynamic results yield solid policy implications. We suggest that governments should not be restrained from promoting inclusive financial policies to a wider segment of the population if short-term benefits are not as significant as they would like to be\textemdash continuous promotion will result in greater effects accumulated over time.

\begin{landscape}
\begin{table}[h!]
    \centering
    \caption{Dynamic DML: How Financial Inclusion Influences Households' Investment Behavior}\label{DDMLresult}
    \def\arraystretch{1.35}    
    \scalebox{0.75}{
   \begin{tabular}{lcccccccccccccccc}
    \toprule
    &\multicolumn{5}{c}{FMP}&\multicolumn{5}{c}{Risky Asset Ratio}&\multicolumn{5}{c}{Sharpe Ratio}\\
    \cmidrule(lr{.75em}){2-6} \cmidrule(lr{.75em}){7-11}\cmidrule(lr{.75em}){12-16}
     &(1)&(2)&(3)&(4)&(5)&(6)&(7)&(8)&(9)&(10)&(11)&(12)&(13)&(14)&(15)\\
      \cline{2-16}
  Panel A.&\multicolumn{15}{l}{Three years dynamic: from 2015 to 2019}\\
         15-19  & 0.005    & -0.004   & 0.033*** & 0.072*** & 0.040*** & 0.045**  & 0.017    & 0.027**  & 0.044*** & 0.023*   & 0.067*** & 0.041**  & 0.021*** & 0.014**  & 0.031**  \\& (0.026)  & (0.010)  & (0.012)  & (0.023)  & (0.014)  & (0.022)  & (0.016)  & (0.011)  & (0.015)  & (0.013)  & (0.017)  & (0.019)  & (0.008)  & (0.006)  & (0.013)  \\ 
         17-19 & -0.049*  & -0.009   & 0.049*** & 0.068*** & 0.037**  & 0.077*** & 0.033**  & 0.039*** & 0.053*** & 0.048*** & 0.104*** & 0.100*** & 0.026*** & 0.013*** & 0.033*** \\& (0.024)  & (0.011)  & (0.011)  & (0.023)  & (0.014)  & (0.021)  & (0.016)  & (0.010)  & (0.013)  & (0.013)  & (0.019)  & (0.019)  & (0.007)  & (0.004)  & (0.011)  \\ 
         19-19 & 0.550*** & 0.351*** & 0.103*** & 0.169*** & 0.095*** & 0.253*** & 0.139*** & 0.088*** & 0.053*** & 0.081*** & 0.142*** & 0.051*** & 0.084*** & 0.027*** & 0.064*** \\& (0.022)  & (0.011)  & (0.011)  & (0.016)  & (0.012)  & (0.017)  & (0.009)  & (0.012)  & (0.010)  & (0.014)  & (0.014)  & (0.005)  & (0.011)  & (0.003)  & (0.013)  \\
          \cline{2-16}    
 Panel B. & \multicolumn{15}{l}{Two years dynamic: 2015 to 2019}\\
     15-19 & -0.021   & -0.012   & 0.057*** & 0.071*** & 0.054*** & 0.086*** & 0.048**  & 0.047*** & 0.043*** & 0.041*** & 0.123*** & 0.135*** & 0.034*** & 0.014*** & 0.044*** \\& (0.026)  & (0.015)  & (0.012)  & (0.023)  & (0.014)  & (0.024)  & (0.022)  & (0.012)  & (0.015)  & (0.014)  & (0.021)  & (0.032)  & (0.009)  & (0.006)  & (0.013)  \\
     19-19 & 0.550*** & 0.351*** & 0.103*** & 0.169*** & 0.095*** & 0.253*** & 0.139*** & 0.088*** & 0.053*** & 0.081*** & 0.142*** & 0.051*** & 0.084*** & 0.027*** & 0.064*** \\& (0.022)  & (0.011)  & (0.011)  & (0.016)  & (0.012)  & (0.017)  & (0.009)  & (0.012)  & (0.010)  & (0.014)  & (0.014)  & (0.005)  & (0.011)  & (0.003)  & (0.013)  \\ 
        \cline{2-16} 
  Panel C. &\multicolumn{15}{l}{Two years dynamic: 2017 to 2019}\\
         17-19 & -0.049** & -0.009   & 0.049*** & 0.068*** & 0.037**  & 0.077*** & 0.033**  & 0.039*** & 0.053*** & 0.048*** & 0.104*** & 0.100*** & 0.026*** & 0.013*** & 0.033**  \\& (0.024)  & (0.011)  & (0.011)  & (0.023)  & (0.014)  & (0.021)  & (0.016)  & (0.010)  & (0.013)  & (0.013)  & (0.019)  & (0.019)  & (0.007)  & (0.004)  & (0.011)  \\ 
         19-19 & 0.550*** & 0.351*** & 0.103*** & 0.169*** & 0.095*** & 0.253*** & 0.139*** & 0.088*** & 0.053*** & 0.081*** & 0.142*** & 0.051*** & 0.084*** & 0.027*** & 0.064*** \\& (0.022)  & (0.011)  & (0.011)  & (0.016)  & (0.012)  & (0.017)  & (0.009)  & (0.012)  & (0.010)  & (0.014)  & (0.014)  & (0.005)  & (0.011)  & (0.004)  & (0.013)  \\
    \bottomrule
    \end{tabular}}
\parbox{0.9\textwidth}{\footnotesize%
\vspace{2eX} Note: Reported are estimated coefficients of the effects of financial inclusion on households' investment behavior using the dynamic DML model. Asterisks indicate ***0.1\%, **1\%, and *5\% significance levels for the $p$-value. Standard errors are reported in the parentheses. We have 17,082 observations in panel A and 11,388 observations in panels B and C. Row names indicate the marginal effect of financial inclusion from year on the year left to the year right to the hyphen. Columns (1), (6), and (11) display the impact of the FA index. Columns (2), (7) and (12) illustrate the influence of whether the household uses digital payment(s). Columns (3), (8), and (13) demonstrate the impact of household ownership of credit card(s). Columns (4), (9), and (14) reflect the effect of whether the household possesses bank account(s). Columns (5), (10) and (15) indicate the influence of whether the household has invested in commercial insurance.}    
\end{table}

\begin{figure}[h!]
    \centering
    \includegraphics[width=\textwidth]{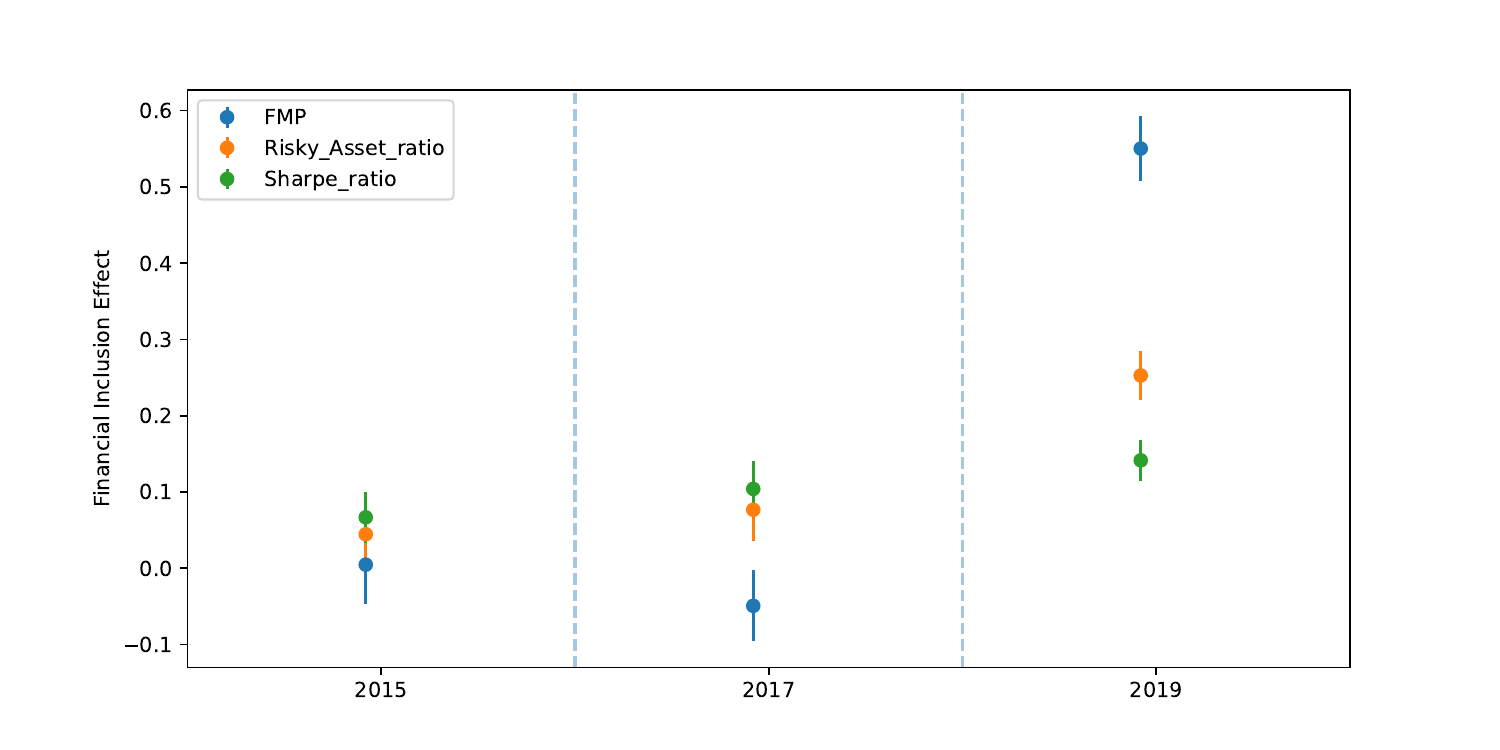}
    \caption{Plot of Coefficients and Error Band for Dynamic DML Method}
    \label{DDML_plot}
\parbox{0.9\textwidth}{\footnotesize%
\vspace{2eX} 
Note: Line plot of coefficients and 95\% confidence intervals estimated by the dynamic DML for the effects of financial inclusion on financial market participation, risky asset ratio, and Sharpe ratio. The financial inclusion service has shown a significant upward trajectory in its impact on these three indicators across three sample years.}  
\end{figure}
\end{landscape}

\section{Further Analyses}\label{heterogeneity}
\subsection{Are Results Robust for Risk-taking Households?}
So far we have identified that there are positive relationships between financial inclusion and households' investment behaviors and have demonstrated an upward trajectory of those effects over time. One particularly noteworthy point is that, as shown in Panel A column (1) of Table \ref{DDMLresult}, we observe insignificant effects in the year 15-19 for financial market participation but not in the year 19-19. We infer that such a time lag in the effects of financial inclusion might be because some households are naturally risk-averse, making it challenging to transition from never investing in risky assets to actively participating in the financial market. To alleviate the concerns of unexpected effects from these households who are naturally risk-averse, we re-run our dynamic DML framework using the subgroup of households who participate in the financial market (FMP=1) at least once in all three years. We aim to investigate the robustness of our main results by comparing the differences between the full sample and the subgroup sample.

Table \ref{DDMLresultFMP=1} presents the results for financial market participants. We observed consistent findings with the main results, revealing a positive and upward trend in coefficients with statistical significance. For example, in column (1) of Table \ref{DDMLresultFMP=1}, the three years dynamics reveal that financial inclusion will lead to an increase in financial market participants' risky asset ratio by 4.8\%, 6.1\%, and 19.3\% for the years 15-19, 17-19, and 19-19, respectively. These results align consistently with the main findings presented in Column (6) of Table \ref{DDMLresult}. Therefore, focusing on financial market participants in our analysis of the dynamic effects of financial inclusion yields consistent trends, reinforcing the robustness of our findings.

\begin{landscape}
\begin{table}[h!]
    \centering
    \caption{Dynamic DML: How Financial Inclusion Influences Households' Investment Behavior(subgroup: FMP=1)}\label{DDMLresultFMP=1}
    \def\arraystretch{1.35}    
    \scalebox{0.75}{
   \begin{tabular}{lcccccccccc}
    \toprule
    &\multicolumn{5}{c}{Risky Asset Ratio}&\multicolumn{5}{c}{Sharpe Ratio}\\
    \cmidrule(lr{.75em}){2-6}\cmidrule(lr{.75em}){7-11}
     &(1)&(2)&(3)&(4)&(5)&(6)&(7)&(8)&(9)&(10)\\
      \cline{2-11}
  Panel A.&\multicolumn{10}{l}{Three years dynamic: from 2015 to 2019}\\
         15-19  &0.048**&0.019&0.022**&0.052***&0.015&0.064***&0.041**&0.020**&0.016**&0.030** \\
                &(0.023)&(0.016)&(0.011)&(0.016)&(0.013)&(0.018)&(0.019)&(0.009)&(0.007)&(0.013) \\
         17-19  &0.061***&0.022&0.026***&0.045***&0.039***&0.108***&0.098***&0.025***&0.012***&0.033*** \\
                &(0.022)&(0.016)&(0.010)&(0.015)&(0.014)&(0.010)&(0.019)&(0.007)&(0.005)&(0.011) \\
         19-19  &0.193***&0.105***&0.068***&0.031***&0.068***&0.138***&0.047***&0.079***&0.028***&0.061*** \\
                &(0.018)&(0.009)&(0.012)&(0.012)&(0.014)&(0.015)&(0.005)&(0.011)&(0.004)&(0.013) \\
          \cline{2-11}    
Panel B. & \multicolumn{10}{l}{Two years dynamic: 2015 to 2019}\\
          15-19&0.081***&0.040*&0.035***&0.052***&0.030**&0.122***&0.132***&0.032***&0.016**&0.042*** \\
              &(0.025)&(0.022)&(0.012)&(0.016)&(0.014)&(0.022)&(0.032)&(0.009)&(0.007)&(0.013) \\
          19-19&0.193***&0.105***&0.068***&0.031***&0.068***&0.138***&0.047***&0.079***&0.028***&0.061*** \\
              &(0.018)&(0.009)&(0.012)&(0.012)&(0.014)&(0.015)&(0.005)&(0.011)&(0.004)&(0.013) \\
        \cline{2-11} 
Panel C. &\multicolumn{10}{l}{Two years dynamic: 2017 to 2019}\\
         17-19&0.061***&0.022&0.026**&0.045***&0.039***&0.108***&0.098***&0.025***&0.012***&0.033*** \\
              &(0.022)&(0.016)&(0.010)&(0.015)&(0.014)&(0.020)&(0.019)&(0.007)&(0.005)& (0.011)\\
         19-19&0.193***&0.105***&0.068***&0.031***&0.068***&0.138***&0.047***&0.079***&0.028***& 0.061***\\
              &(0.018)&(0.009)&(0.012)&(0.012)&(0.014)&(0.015)&(0.005)&(0.011)&(0.004)&(0.013) \\
        
    \bottomrule
    \end{tabular}}
\parbox{0.9\textwidth}{\footnotesize%
\vspace{2eX} Note: Reported are estimated coefficients of the effects of financial inclusion on households' investment behavior using the dynamic DML model. Asterisks indicate ***0.1\%, **1\%, and *5\% significance levels for the $p$-value. Standard errors are reported in the parentheses. We selected individuals who participated in the financial market at least once for all three years, resulting in 5,162 households. Consequently, we have 15,486 observations in panel A and 10,324 observations in panels B and C. Row names indicate the marginal effect of financial inclusion from year on the year left to the year right to the hyphen. Columns (1), (6), and (11) display the impact of the FA index. Columns (2), (7) and (12) illustrate the influence of whether the household uses digital payment(s). Columns (3), (8), and (13) demonstrate the impact of household ownership of credit card(s). Columns (4), (9), and (14) reflect the effect of whether the household possesses bank account(s). Columns (5), (10) and (15) indicate the influence of whether the household has invested in commercial insurance.}    
\end{table}
\end{landscape}

\subsection{Alternative Measurement of Households Financial Inclusion}
As a second robustness check, we address the potential concerns arising from weight selection in our measure of household financial inclusion. Although there is no standard procedure for assigning weights over dimensions and indicators when constructing an individual-level financial inclusion index, we opt for the average of the four individual financial inclusion measures, largely due to the concerns regarding model interpretability. Following \cite{cheng2023does}, we alternatively consider using the entropy method to weight the financial inclusion index for households and call it FA score. One of the advantages is that it overcomes randomness-related issues and effectively solves information overlap between multi-index variables. 

Table \ref{factored_FA_FE} presents the coefficient estimates of the fixed effect model, using the household-level financial index calculated under the weights determined by the entropy method. We find that the effects of financial inclusion on financial market participation, risky asset ratio, and Sharpe ratio continue to be highly significant in statistical and economic terms once we control for household characteristics and other demographic factors. Moving to DML and dynamic DML frameworks, the estimated average treatment and marginal effects are consistent with our main results in Table \ref{DMLresult} and Table \ref{DDMLresult}. The pair-wise differences between estimated coefficients for our baseline results in Table \ref{FEresult}, Table \ref{DMLresult}, Table \ref{DDMLresult} and robustness results are indeed quite small, suggesting that our baseline results are robust to a potential bias resulting from inappropriate weights selection in the measure of household financial inclusion.

\begin{table}[h!]
\centering 
    \caption{Robustness: FA Score as an Alternative Measure of Financial Inclusion}\label{robustFAscore1}
\scalebox{0.8}{
\begin{tabular}{lcccccc}
  \toprule
 &\multicolumn{2}{c}{FMP}&\multicolumn{2}{c}{Risky Asset Ratio}&  \multicolumn{2}{c}{Sharpe Ratio}\\
    \cmidrule(lr{.75em}){2-3} \cmidrule(lr{.75em}){4-5}\cmidrule(lr{.75em}){6-7}
     & (1) & (2) & (3) & (4) &   (5) &(6)\\ 
  \cmidrule{2-7}
(Intercept) & 0.467***&  0.232***&0.154***& 0.076***&  0.415***&0.356***\\
                &(0.004)&(0.036)& (0.003)&(0.022)&  (0.001)&(0.012)\\
FA\_score & 0.747***& 0.406***& 0.304***&0.188***&  0.147***&0.115*** \\
            &(0.016)& (0.018)& (0.010)&(0.011)&  (0.005)&(0.006)\\
age               &&    -0.002***&&0.000&  &0.000***   \\
                    &&    (0.000)&&(0.000)&  &(0.000)   \\
male             &&     0.011&&0.001&  &0.001    \\
                  &&     (0.009)&&(0.006)&  &(0.003)    \\
marriage        &&       0.017&&-0.005&  &-0.002     \\
                  &&      (0.011)&&(0.007)&  &(0.004)   \\
ind\_commer\       &&      0.042***&&0.011*&  &-0.006* \\  
                  &&      (0.010)&&(0.006)&  &(0.003)   \\
edu                &&      0.035***&&0.015&  &0.015***    \\
                     &&     (0.003)&&(0.002)&  &(0.001)   \\
health              &&       -0.009**&&-0.002&  &0.003***   \\
                 &&       (0.004)&&(0.002)&  &(0.001)    \\
oldsum            &&        -0.003&&0.002&  &0.006***    \\
                   &&      (0.005)&&(0.003)&  &(0.002)    \\
youngsum           &&        -0.027***&&-0.001&  &0.003   \\
                     &&      (0.006)&&(0.004)&  &(0.002)   \\
family\_size           &&     0.023***&&0.003&  &-0.003***  \\
                      &&      (0.003)&&(0.002)&  &(0.001)\\
high\_asset          &&      -0.025**&&0.029&  &0.036***   \\
                           &&(0.011)&&(0.007)&  &(0.003)   \\
middle\_asset          &&    0.073***&&0.055***&  &0.018***   \\
                        &&     (0.009)&&(0.005)&  &(0.003)    \\
fina\_know              &&   0.024**&&0.025***&  &0.025***  \\
                       &&      (0.012)&&(0.007)&  &(0.004)    \\
risk\_prefer            &&    0.035***&&0.045***&  &0.066***   \\
                            &&   (0.013)&&(0.008)&  &(0.004)   \\
east                       &&  0.028&&0.047***&  &0.047***   \\
                         &&   (0.027)&&(0.017)&  &(0.009)    \\
middle                     &&     0.074***&&0.028*&  &0.007   \\
                            &&   (0.025)&&(0.016)&  &(0.008)    \\
rural                       && -0.032*&&-0.032***&  &-0.012***\\
                            && (0.008)&&(0.005)&  &(0.003)\\
year2017                   && 0.208***&&0.034*** &  &-0.024***\\
                           && (0.009)&&(0.006)&  &(0.003)\\
year2019                  && 0.336***&&0.034***&  &-0.065***\\
                          && (0.009)&&(0.006)&  &(0.003)\\
                              &&&&&  &\\
Obs. & 17,082& 17,082& 17,082& 17,082&  17,082 &17,082\\ 
Province FE & No&Yes&No&Yes&  No&Yes\\
\bottomrule
\end{tabular}}
\label{factored_FA_FE}
\parbox{0.9\textwidth}{\footnotesize%
\vspace{2eX} Note: Reported are fixed effect OLS estimated coefficients of the effects of financial inclusion on households' investment behavior using FA score. The FA score is calculated by employing the entropy method weighting on the four indicators of financial inclusion. Asterisks indicate ***0.1\%, **1\%, and *5\% significance levels for the $p$-value. Standard errors are reported in the parentheses.} 
\end{table}

\begin{table}[h!]
    \centering 
    \caption{Robustness: FA Score as an Alternative Measure of Financial Inclusion}\label{robustFAscore2}
    \def\arraystretch{1.5}
    \scalebox{0.8}{
    \begin{tabular}{lccc}
    \toprule
& FMP & Risky Asset Ratio &Sharpe Ratio \\
  \cmidrule{2-4}
   Panel A  &  \multicolumn{3}{l}{Three year cross sectional analysis (DML)}\\
         15-15 &0.511***&0.266***&0.094***\\ 
               &(0.035)&(0.022)&(0.014)\\ 
         17-17&0.264***&0.169***&0.117***\\  
        &(0.029)&(0.023)&(0.016)\\ 
         19-19&0.372***&0.193***&0.122***\\ 
           &(0.020)&(0.020)&(0.015)\\  
  Panel B  & \multicolumn{3}{l}{Three years dynamic (DDML)}\\
         15-19 &-0.006 &0.019&0.049***\\  
               &(0.019) &(0.019)&(0.016)\\  
         17-19&-0.059*** &0.043**&0.077***\\  
         &(0.019) &(0.018)&(0.016)\\   
         19-19&0.464*** &0.243***&0.147***\\   
           &(0.019) &(0.017)&(0.016)\\       
        \bottomrule
    \end{tabular}}
\parbox{0.9\textwidth}{\footnotesize%
\vspace{2eX} Note: Reported are DML estimated coefficients of the effects of financial inclusion on households' investment behavior using FA score. The FA score is calculated by employing the entropy method weighting on the four indicators of financial inclusion. Asterisks indicate ***0.1\%, **1\%, and *5\% significance levels for the $p$-value. Standard errors are reported in the parentheses.}   
\end{table}

\subsection{City Level Financial Inclusion Status as an Instrument}
To further enhance the robustness of our analysis and address potential endogeneity concerns in our investigation of the effects of financial inclusion on China's household willingness to participate in the financial market, asset allocation and investment efficiency, we turn to an instrumental variable (IV) approach. Our instrumental variable, the city average of the financial inclusion index, is selected with careful consideration. We posit that a city-level index is a suitable instrument because it is plausibly exogenous to the household-level asset allocation and investment efficiency decisions, yet it is strongly correlated with the overall level of financial inclusion experienced by households within the city. This instrument is expected to satisfy the two key instrumental variable assumptions: relevance and exogeneity. Firstly, we argue that the city-level financial inclusion index is highly relevant to our research question since it captures the broader financial environment within which households make their financial decisions, yet it is unlikely to directly influence their specific portfolio choices or investment efficiency. Secondly, we maintain that the city-level index is exogenous, as it is determined by local financial policy and infrastructure, and it is unlikely to be affected by the specific decisions of individual households within the city. Through this instrumental variable strategy, we aim to provide additional confidence in the causal relationship between financial inclusion and household portfolio allocation and investment efficiency, strengthening the robustness of our findings.

We report results from fixed effect OLS+IV in Table \ref{IV_FE} and from DML+IV in Table \ref{DML_IV} using the city-level index as the instrument described above. The effects of financial inclusion on financial market participation, risky asset ratio, and Sharpe ratio are still statistically significant for both OLS+IV and DML+IV approaches, further validating our main analysis. One particularly noteworthy point is there are observed large differences in estimated coefficients between baseline fixed effect OLS and fixed effect OLS+IV approach, while the differences between baseline DML and DML+IV are negligible. For instance, the estimated coefficient for financial market participation in fixed effect OLS+IV is 83\% (column (1) of Table \ref{FEresult}, almost twice as much as that of in baseline fixed effect OLS (column (1) of Table \ref{FEIV}. In theory, the IV approach is designed to address endogeneity issues that may exist in the baseline OLS model, whereas DML incorporates non-linear and non-parametric functional forms to extract unconfounded signals to obtain an unbiased estimate. We find evidence that estimated coefficients are more stable regardless of whether we include the instrument or not in DML, but coefficients vary significantly if we add IV to fixed effect OLS. This finding further validates the soundness of empirical strategy\textemdash DML provides more consistent estimates and less biased policy implications, whereas the fixed effect model is subject to a precise selection of control variables.

\begin{table}[h!]
\centering 
    \caption{Robustness (FE+IV): City Level Financial Inclusion Status as a Proxy for a Household's Peers}\label{FEIV}
\scalebox{0.8}{
\begin{tabular}{lcccccc}
  \toprule

 &\multicolumn{2}{c}{FMP}&\multicolumn{2}{c}{Risky Asset Ratio}&  \multicolumn{2}{c}{Sharpe Ratio}\\
    \cmidrule(lr{.75em}){2-3} \cmidrule(lr{.75em}){4-5}\cmidrule(lr{.75em}){6-7}
     & (1) & (2) & (3) & (4) &   (5) &(6)\\ 
  \cmidrule{2-7}
(Intercept) & -0.118***&  0.002&-0.049***& -0.122***&  0.371***&0.252***\\
                      &(0.020)&(0.061)& (0.011)&(0.040)&  (0.006)&(0.020)\\
  FA\_ratio & 2.095***& 0.831***& 0.748***&0.626***&  0.198***&0.337*** \\
                   &(0.058)& (0.131)& (0.033)&(0.085)&  (0.017)&(0.044)\\
age               &&    -0.000***&&0.001***&  &0.001***   \\
                    &&    (0.001)&&(0.000)&  &(0.000)   \\
male             &&     0.014&&0.006&  &0.003    \\
                  &&     (0.009)&&(0.006)&  &(0.003)    \\
marriage        &&       0.014&&-0.006&  &-0.002     \\
                  &&      (0.011)&&(0.008)&  &(0.004)   \\
ind\_commer\       &&      0.029**&&-0.004&  &-0.014*** \\  
                  &&      (0.012)&&(0.008)&  &(0.004)   \\
edu                &&      0.023***&&0.003&  &0.009***    \\
                     &&     (0.005)&&(0.003)&  &(0.002)   \\
health              &&       -0.005**&&0.002&  &0.005***   \\
                 &&       (0.004)&&(0.002)&  &(0.001)    \\
oldsum            &&        0.002&&0.007**&  &0.008***    \\
                   &&      (0.005)&&(0.003)&  &(0.002)    \\
youngsum           &&        -0.023***&&0.002&  &0.004   \\
                     &&      (0.007)&&(0.004)&  &(0.002)   \\
family\_size           &&     0.017***&&-0.003&  &-0.006***  \\
                      &&      (0.003)&&(0.002)&  &(0.001)\\
high\_asset          &&      -0.047***&&0.007&  &0.024***   \\
                           &&(0.013)&&(0.008)&  &(0.004)   \\
middle\_asset          &&    0.041***&&0.023***&  &0.002   \\
                        &&     (0.013)&&(0.008)&  &(0.004)    \\
fina\_know              &&   0.001&&0.001&  &0.013***  \\
                       &&      (0.013)&&(0.008)&  &(0.005)    \\
risk\_prefer            &&    0.019&&0.027***&  &0.056***   \\
                            &&   (0.014)&&(0.009)&  &(0.005)   \\
east                       &&  0.018&&0.035**&  &0.041***   \\
                         &&   (0.027)&&(0.017)&  &(0.009)    \\
middle                     &&     0.088***&&0.042**&  &0.014*    \\
                            &&   (0.026)&&(0.017)&  &(0.008)    \\
rural                       && -0.016*&&-0.016***&  &-0.005***\\
                            && (0.008)&&(0.006)&  &(0.003)\\
year2017                   && 0.204***&&0.030*** &  &-0.026***\\
                           && (0.009)&&(0.006)&  &(0.003)\\
year2019                  && 0.308***&&-0.000&  &-0.082***\\
                          && (0.014)&&(0.009)&  &(0.005)\\
                              &&&&&  &\\
Obs. & 17,082& 17,082& 17,082& 17,082&  17,082 &17,082\\ 
Province FE & No&Yes&No&Yes&  No&Yes\\
\bottomrule
\end{tabular}}
\label{IV_FE}
\parbox{0.9\textwidth}{\footnotesize%
\vspace{2eX} Note: Reported are estimated coefficients of the effects of financial inclusion on households' investment behavior using fixed effect OLS+IV approach. We use city-level financial inclusion status as an instrument variable and apply 2SLS method in estimating the individual financial inclusion effect on household financial market participation, and its efficiency. Asterisks indicate ***0.1\%, **1\%, and *5\% significance levels for the $p$-value. Standard errors are reported in the parentheses.} 
\end{table}

\begin{table}[h!]
    \centering 
    \caption{Robustness (DML+IV): Using City Level Financial Inclusion Status as a Proxy for a Household's Peers}\label{robustIV}
    \def\arraystretch{1.5}
    \scalebox{0.8}{
    \begin{tabular}{lccc}
    \toprule
& FMP & Risky Asset Ratio &Sharpe Ratio \\
  \cmidrule{2-4}
   Panel. &  \multicolumn{3}{l}{Three Year Cross-sectional Analysis (DML)}\\
         15-15 &0.524***&0.276***&0.122***\\ 
               &(0.043)&(0.026)&(0.016)\\ 
         17-17&0.363***&0.188***&0.143***\\  
        &(0.036)&(0.026)&(0.017)\\ 
         19-19&0.461***&0.191***&0.117***\\ 
           &(0.024)&(0.020)&(0.013)\\   
        \bottomrule
    \end{tabular}}
    \label{DML_IV}
\parbox{0.9\textwidth}{\footnotesize%
\vspace{2eX} Note: Reported are estimated coefficients of the effects of financial inclusion on households' investment behavior using DML+IV approach. We use city-level financial inclusion status as an instrument variable and apply 2SLS method in estimating the individual financial inclusion effect on household financial market participation, and its efficiency. Asterisks indicate ***0.1\%, **1\%, and *5\% significance levels for the $p$-value. Standard errors are reported in the parentheses.}    
\end{table}

\subsection{Individual Heterogeneity}
Prior literature suggests that risk-taking behavior and investment efficiency can be influenced by household head's characteristics (e.g. \cite{campbell2006household};\cite{guiso2013household};\cite{choi2020matters};\cite{gomes2021household}). Households with pessimistic characters, for instance, better education in financial literacy and greater asset levels, are more willing to participate in the equity market compared to optimistic persons. To check if our main results are heterogeneous by differences in households' characters, we perform heterogeneity analysis to uncover the effectiveness of particular factors in portfolio diversification, providing more comprehensive insights into the potential benefits of improved financial inclusion for different segments of sample populations. Since fixed effect models, DML and dynamic DML all provide positive results, we only report coefficients estimated by the dynamic DML method in heterogeneity analysis due to its ability to show time dynamics of financial inclusion. We consider three types of individual heterogeneity: (i) whether households live in rural regions or not, (ii) households' educational level, and (iii) households' asset level.

\begin{figure}[h!]
    \centering
    \includegraphics[width=\textwidth]{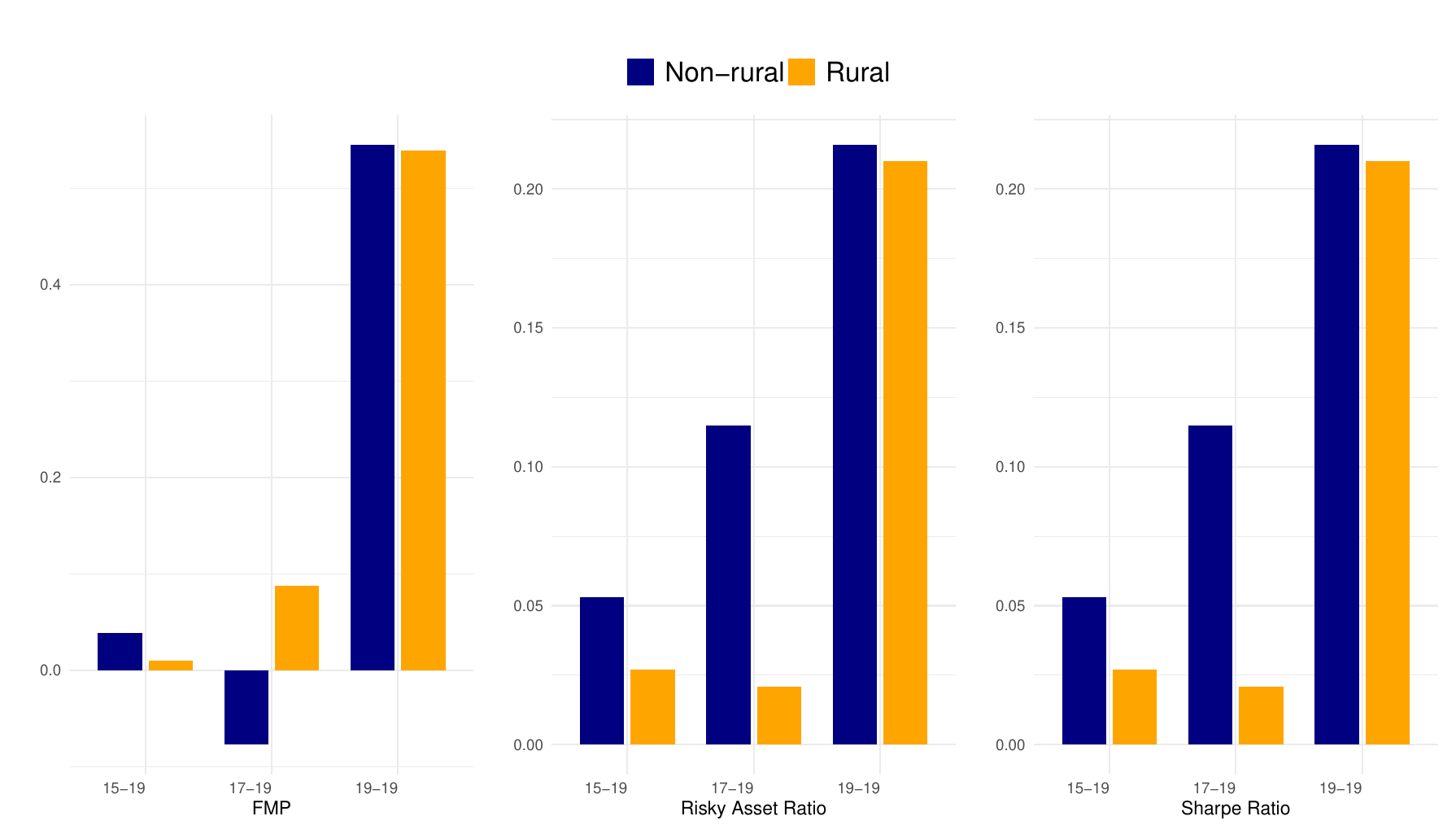}
    \caption{Heterogeneity Plot: Rural vs. Non-Rural}
    \label{Fig_rural}
\parbox{0.9\textwidth}{\footnotesize%
\vspace{2eX} 
Note: barplot of coefficients estimated by the dynamic DML for the heterogeneity effects of financial inclusion on financial market participation, risky asset ratio, and Sharpe ratio between rural and non-rural households. The financial inclusion service has shown a significant upward trajectory in its impact on these three indicators across three sample years. Non-rural households are colored in navy blue, whereas rural households are colored in orange. Coefficients are grouped according to the measure of household investment behavior.}      
\end{figure}

\textbf{Do rural residents benefit more?} We first explore whether better financial inclusion access yields heterogeneous effects on having rural residency or not. Table \ref{Hetero_Rural} and Figure \ref{Fig_rural} present the Dynamic DML modeling results on rural and non-rural residents. Although residency only subtly affects the allocation between risk-free and risk assets, urban residents tend to benefit greatly from having better financial services than households living in rural areas in terms of investment efficiency and portfolio diversification. In most circumstances, the effects are several times better. For instance, the 17-19 effect for non-rural residents is 0.115 (\textit{t-stat=0.025}), four times larger than the contemporary effect for rural residents. Compared with households living in rural areas, urban residents often have better access to financial education and information due to the presence of financial institutions, educational institutions, and financial literacy programs \citep{gaudecker2015does,yang2022fintech,yang2023digital}. In addition, urban residents often have better knowledge about using advanced technology and financial inclusion services, which would likely result in greater effects on investment efficiency. In contrast, the difference between the effects of better financial inclusion on financial market participation for rural residents and non-rural residents is negligible, reaching 53.9\% and 54.5\% of 19-19 effects for rural residents and non-rural residents, respectively.

\begin{figure}[h!]
    \centering
    \includegraphics[width=\textwidth]{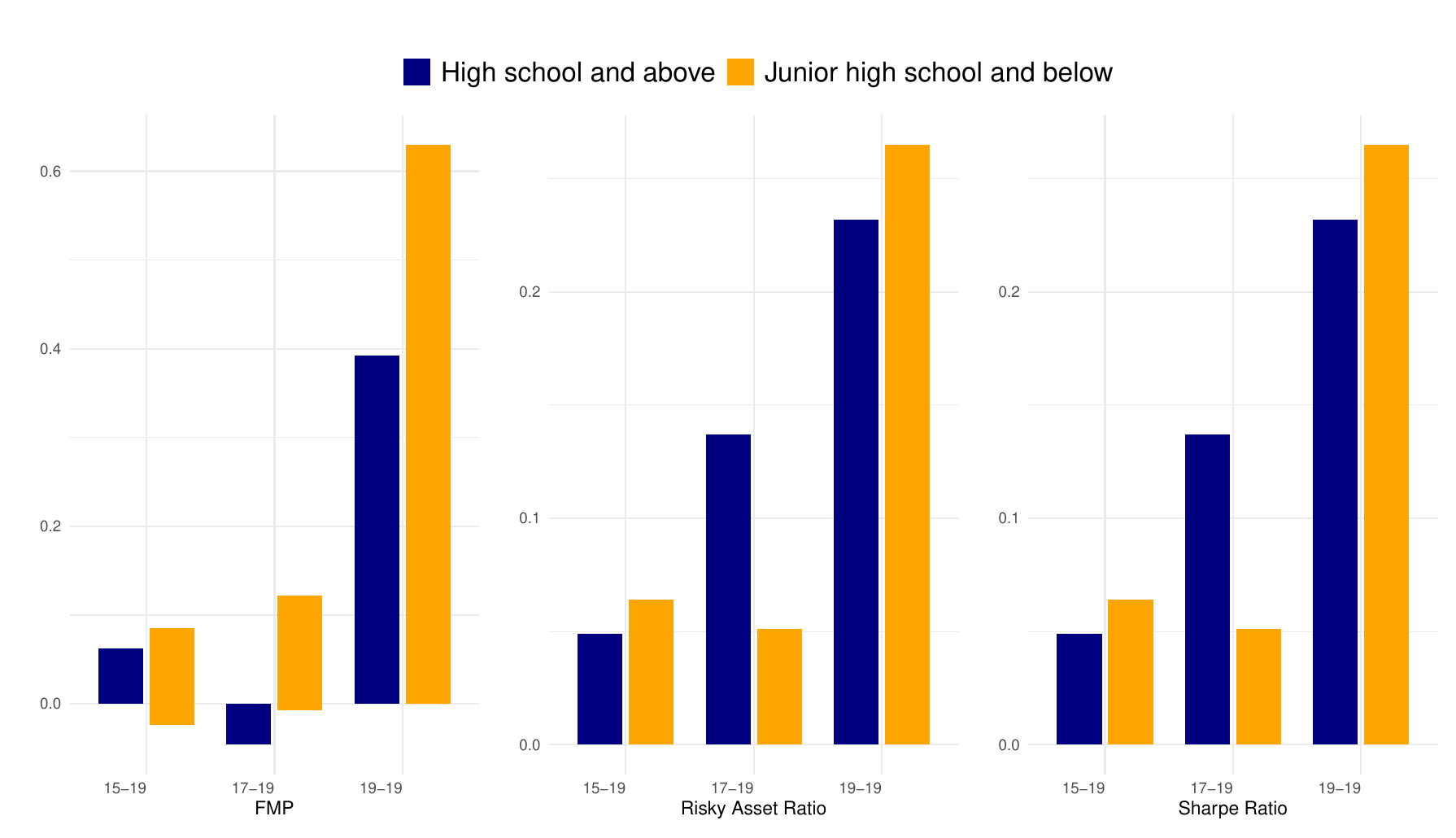}
    \caption{Heterogeneity Plot: High School and Above vs. Junior High School and Below}
        \label{Fig_school}
\parbox{0.9\textwidth}{\footnotesize%
\vspace{2eX} 
Note: barplot of coefficients estimated by the dynamic DML for the heterogeneity effects of financial inclusion on financial market participation, risky asset ratio, and Sharpe ratio between households with high school and above education and junior high school and below education backgrounds. The financial inclusion service has shown a significant upward trajectory in its impact on these three indicators across three sample years. High school and above-educated households are colored in navy blue, whereas junior high school and below-educated households are colored in orange. Coefficients are grouped according to the measure of household investment behavior.}      
\end{figure}

\textbf{The better educated the greater effects?} Turning to the dynamic effects of financial inclusion among people with different educational backgrounds presented in Table \ref{Hetero_Edu} and Figure \ref{Fig_school}, we find marginal effects on the financial market participation (as shown in columns (1) and (4)) are large in years 19-19 and are small for years 15-19 and 17-19. Effects are particularly pronounced for households with junior high school and below education, indicating less educated households are more willing to participate in the financial market than better-educated households under the effect of financial inclusion. With regards to the risky asset ratio and Sharpe ratio, a glance reveals that estimated coefficients are all significantly positive, except for 15-19 marginal effects for households with high school and above education background (as shown in columns (5) and (6)). The effects of financial inclusion penetration on investment efficiency are more pronounced for households with high school and above education, while financial inclusion effects for people with lower education levels are more subtle. The level of financial literacy, critical thinking skills, and access to resources are possible explanations for better utilization of financial inclusion services, leading to a better ability to make informed financial decisions, manage risk, and invest efficiently.

\begin{figure}[h!]
    \centering
    \includegraphics[width=\textwidth]{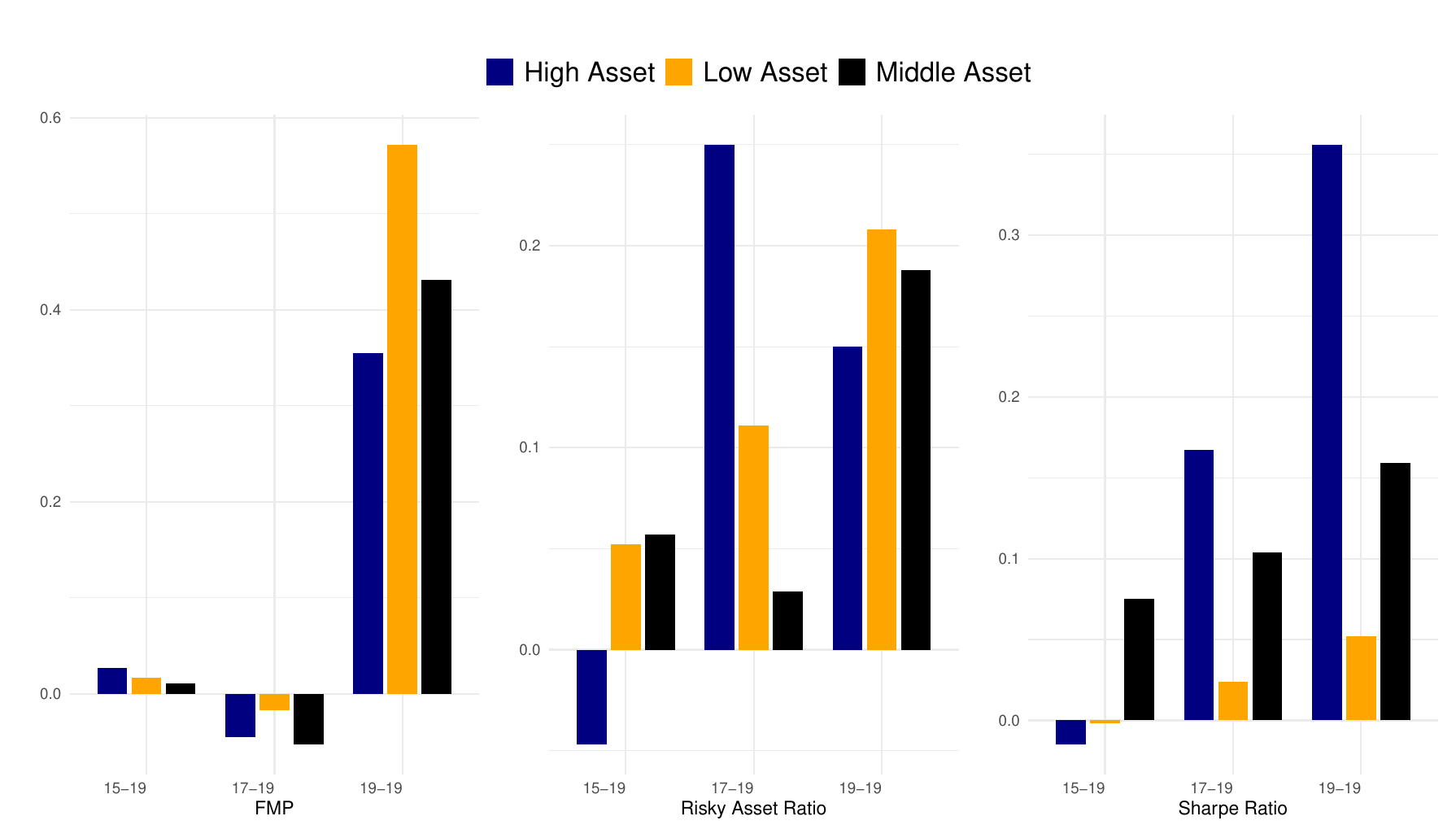}
    \caption{Heterogeneity Plot: Low, Middle and High Asset}
        \label{Fig_asset}
\parbox{0.9\textwidth}{\footnotesize%
\vspace{2eX} 
Note: barplot of coefficients estimated by the dynamic DML for the heterogeneity effects of financial inclusion on financial market participation, risky asset ratio, and Sharpe ratio among households with high, middle and low asset sizes. The financial inclusion service has shown a significant upward trajectory in its impact on these three indicators across three sample years. High-asset households are colored in navy blue, middle-asset households are colored in black, and low-asset households are colored in orange. Coefficients are grouped according to the measure of household investment behavior.}  
\end{figure}

\textbf{Does asset level matter?} Regarding the asset levels of households, we categorize households with assets greater than 3 million Yuan as the high asset level group, households with assets between half a million Yuan and 3 million Yuan as the middle asset level group, and households with assets less than half a million Yuan as the low asset level group.\footnote{Since households may change between groups in different years, our analysis separates households by their 2019 asset levels.} Table \ref{Hetero_asset} and Figure \ref{Fig_asset} illustrate the dynamic effects of heterogeneity across the three different household asset levels. Households with higher asset levels benefit more from financial inclusion in terms of the Sharpe ratio, whereas households with low assets are more likely to participate in the financial markets following the dynamic effects of financial inclusion. For example, considering the marginal effect of financial inclusion on the Sharpe ratio between 2017 and 2019, high asset-level households experience a substantial impact, with a coefficient of 0.163 (\textit{t-stat} = 2.063), whereas their low asset-level counterparts exhibit a much smaller effect at 0.024 (\textit{t-stat} = 2.400). This disparity persists in the 2019-2019 marginal effects on the Sharpe ratio, where the high-level asset group demonstrates a coefficient of 0.356 (t-stat = 4.238), compared to the modest 0.052 (\textit{t-stat} = 5.778) observed in the low asset group. In contrast, the financial market participation in 19-19 effects are 57.2\% for the low asset class and 35.5\% for the high asset class. The correlation between higher financial asset levels and a more pronounced impact from financial inclusion is not coincidental. Households with greater financial assets often possess higher levels of education, greater financial literacy, and increased access to other social resources. Moreover, their ability to adapt swiftly to emerging opportunities contributes to their enhanced capacity to benefit from the penetration of financial inclusion.

\textbf{Upward or downward trajectory?} In addition to observing significant heterogeneous effects of financial inclusion on households' willingness to participate in the financial markets, risk-taking behavior and investment efficiency, we also find the dynamics within each heterogeneous group generally exhibit an upward trajectory, which is consistent with our results in Table \ref{DDMLresult}. We notice an interesting finding: the extent of household benefit derived from financial inclusion appears to be contingent on their residency, educational level and asset level. Rural residents, junior high school and below and middle asset level groups seem to be affected by a longer time period whereas the urban, high school and above, low and high assets groups are affected only for two periods within our data framework. 

\begin{table}[h!]
    \centering 
    \caption{Heterogeneity: Rural and Non-Rural}\label{Hetero_Rural}
    \def\arraystretch{1.5}
    \scalebox{0.8}{
    \begin{tabular}{ccccccc}
    \toprule
    &  &\multicolumn{2}{c}{Rural}&  &\multicolumn{2}{c}{Non-Rural}\\
     &   FMP&Risky Asset Ratio&Sharpe Ratio &   FMP&Risky Asset Ratio&Sharpe Ratio \\
     &  (1)&(2)&(3)&  (4)&(5)&(6)\\
      \cline{2-7}
         15-19 &  0.008&0.010 &0.027&  0.031&0.039&0.053* \\ 
               &  (0.073)&(0.044) &(0.017)&  (0.028)&(0.026) &(0.021) \\ 
         17-19&  0.088&0.074** &0.021*&   -0.077***&0.090*** &0.115***\\ 
            &  (0.058)&(0.036) &(0.012)&  (0.025&(0.026) &(0.024) \\ 
         19-19&  0.539***&0.210*** &0.049***&  0.545***&0.216*** &0.195*** \\ 
              &  (0.038)&(0.024) &(0.008)&  (0.028)&(0.024) &(0.021) \\ 
        \emph{No. obs}&  6,732&6,732&6,732&  10,350&10,350&10,350\\
        \bottomrule
    \end{tabular}}
\parbox{0.9\textwidth}{\footnotesize%
\vspace{2eX} Note: Asterisks indicate ***1\%, **5\%, and *10\% significance levels for the $p$-value. Standard errors are reported in the parentheses. We separate the data based on whether the households reside in rural or non-rural areas in the year 2019. In particular, if a household was situated in a rural area during the year 2019, then the household is classified as part of the Rural subset, regardless of the family's location in 2015 or 2017.}    
\end{table}

\begin{table}[h!]
    \centering 
    \caption{Heterogeneity: Education Level}\label{Hetero_Edu}
    \def\arraystretch{1.5}
    \scalebox{0.8}{
    \begin{tabular}{ccccccc}
    \toprule
    &  &\multicolumn{2}{c}{Junior High School and Below}& &\multicolumn{2}{c}{High School and Above}\\
     &   FMP&Risky Asset Ratio&Sharpe Ratio &  FMP&Risky Asset Ratio& Sharpe Ratio \\
     &  (1)&(2)&(3)& (4)&(5)&(6)\\
      \cline{2-7}
         15-19 &  -0.024&0.085**&0.064***& 0.062***&0.025&0.049\\
               &  (0.049)&(0.035)&(0.016)& (0.029)&(0.033)&(0.030)\\
         17-19 &  -0.007&0.122***&0.051***& -0.046*&0.063*&0.137***\\
            &  (0.043)&(0.030)&(0.014)& (0.026)&(0.033)&(0.032)\\
         19-19&  0.630***&0.265***&0.087***& 0.392***&0.157***&0.232***\\
              &  (0.029)&(0.021)&(0.011)& (0.034)&(0.034)&(0.031)\\
        \emph{No. obs}&   11,514&11,514&11,514& 5,568&5,568&5,568\\
        \bottomrule
    \end{tabular}}
\parbox{0.9\textwidth}{\footnotesize%
\vspace{2eX} Note: Asterisks indicate ***1\%, **5\%, and *10\% significance levels for the $p$-value. Standard errors are reported in the parentheses. We separate the data based on the education levels of the household head in the year 2019.}    
\end{table}

\begin{table}[h!]
    \centering 
    \caption{Heterogeneity: Asset Level}\label{Hetero_asset}
    \def\arraystretch{1.5}
    \scalebox{0.6}{
    \begin{tabular}{cccccccccc}
    \toprule
    &  &\multicolumn{2}{c}{Low}& &\multicolumn{2}{c}{Middle}& &\multicolumn{2}{c}{High}\\
     &   FMP&Risky Asset Ratio&Sharpe Ratio &  FMP&Risky Asset Ratio& Sharpe Ratio &  FMP&Risky Asset Ratio& Sharpe Ratio \\
      &   (1)&(2)&(3)&  (4)&(5)& (6)&  (7)&(8)& (9)\\
      \cline{2-10}
      15-19 &  0.017&0.052 &-0.002 & 0.011&0.057* &0.075*** & 0.027&-0.047&-0.015\\ 
            &  (0.062)&(0.041) &(0.014) & (0.031)&(0.033) &(0.029) & (0.054)&(0.063) &(0.052) \\ 
    17-19 &  -0.017&0.111*** &0.024** & -0.053*&0.029 &0.104*** & -0.045&0.250*** &0.167**\\  
        &  (0.053)&(0.035) &(0.010) & (0.026)&(0.034) &(0.029) & (0.063)&(0.064) &(0.079) \\ 
    19-19&  0.572***&0.208*** &0.052*** & 0.431***&0.188*** &0.159*** & 0.355***&0.150** &0.356*** \\ 
        &  (0.035)&(0.024) &(0.009) & (0.034)&(0.033) &(0.024) & (0.072)&(0.068)&(0.084) \\  
        \emph{No. obs}&  9,426&9,426&9,426& 6,345&6,345&6,345& 1,311&1,311&1,311\\
        \bottomrule
    \end{tabular}}
\parbox{0.9\textwidth}{\footnotesize%
\vspace{2eX} Note: Asterisks indicate ***1\%, **5\%, and *10\% significance levels for the $p$-value. Standard errors are reported in parentheses. We separate the data by asset levels of households in the year of 2019.}    
\end{table}

\section{Conclusion}\label{Conclusion}
In recognition of the importance of the effects of the promotion of financial inclusion on household investment behavior and efficiency as a crucial development strategy for better financial development, this paper uses survey data from the China Household Finance Survey to investigate whether having better access to financial inclusion services encourage households to participate in the financial market, increases the diversification of household equity investments as well as the efficiency of portfolio. Our results from fixed effect and double machine learning both highlight the profound significance of positive financial inclusion effects on financial market participation, household financial asset allocation, and investment efficiency, offering compelling evidence for the imperative implementation of financial inclusion-related policies. Our dynamic analysis further strengthens this conclusion by revealing consistent upward trajectories in household financial well-being resulting from the provision of financial inclusion services, thus affirming our hypothesis that the gradual penetration of such services exerts an increasingly substantial impact on the financial well-being of households. Our results are robust to include IV in model specification and alternative measurement of household financial inclusion. Individual heterogeneity presents in the sense that urban residents benefit more than rural residents, the better educated the greater effects, and asset level matters. 

Our findings shed light on the pivotal role of financial inclusion in reshaping household financial landscapes and highlight the necessity for inclusive policies that bridge existing divides, thereby fostering enduring financial well-being for all members of society. From a policy perspective, two critical insights emerge. Firstly, the gradual expansion of financial inclusion services not only extends benefits to a broader spectrum of individuals but also fosters sustainability in the pursuit of enhanced financial well-being. Secondly, it is essential to acknowledge the existence of a digital divide that persists among various educational and socioeconomic groups. Policymakers should prioritize strategies aimed at narrowing these disparities, ensuring that the advantages of financial inclusion are accessible and equally beneficial to all, regardless of their educational or economic backgrounds.

As we continue to delve deeper into the advantages of financial inclusion on household investment behavior, the potential for significant positive social and economic impacts becomes increasingly evident. Future research avenues may involve more comprehensive assessments of household portfolios and household characteristics. One promising direction could involve combining administrative data and measures of portfolio diversification drawn from household surveys. This approach would yield more precise and quantitatively meaningful diversification metrics for respondents, as opposed to relying solely on index returns as proxies for households in our study.

\bibliographystyle{elsarticle-harv}\biboptions{authoryear} 
\bibliography{reference}
\end{document}